\documentclass[twocolumn,prb,superscriptaddress,aps,10pt,bibnotes,amsfonts,amsmath,amssymb,floatfix,nolongbibliography]{revtex4-2}
\usepackage{color}
\usepackage{soul}
\usepackage{hhline}	
\usepackage{graphicx}
\usepackage{color}
\usepackage{dcolumn}
\usepackage{multirow}
\usepackage{booktabs}
\usepackage{afterpage}
\usepackage{amsmath,cases}
\usepackage{braket}
\usepackage{array}
\usepackage{placeins}
\renewcommand{\bm}[1]{\boldsymbol{#1}}
\usepackage{newtxtext}
\usepackage[varg,upint]{newtxmath}

\usepackage{hyperref}
\hypersetup{
 setpagesize=false,
 bookmarksnumbered=true,%
 bookmarksopen=true,%
 colorlinks=true,%
 linkcolor=blue,
 citecolor=blue,
 urlcolor=blue,
}
\usepackage{soul,xcolor}

\newcommand{\im}{\mathrm{i}}
\newcommand{\e}{\mathrm{e}}

\makeatletter
\DeclareSymbolFont{cmlargesymbols}{OMX}{cmex}{m}{n}
\let\sum\relax
\DeclareMathSymbol{\sum}{\mathop}{cmlargesymbols}{"50}
\DeclareFontFamily{OMX}{MnSymbolE}{}
\DeclareSymbolFont{MnLargeSymbols}{OMX}{MnSymbolE}{m}{n}
\SetSymbolFont{MnLargeSymbols}{bold}{OMX}{MnSymbolE}{b}{n}
\DeclareFontShape{OMX}{MnSymbolE}{m}{n}{
    <-6>  MnSymbolE5
   <6-7>  MnSymbolE6
   <7-8>  MnSymbolE7
   <8-9>  MnSymbolE8
   <9-10> MnSymbolE9
  <10-12> MnSymbolE10
  <12->   MnSymbolE12
}{}
\DeclareFontShape{OMX}{MnSymbolE}{b}{n}{
    <-6>  MnSymbolE-Bold5
   <6-7>  MnSymbolE-Bold6
   <7-8>  MnSymbolE-Bold7
   <8-9>  MnSymbolE-Bold8
   <9-10> MnSymbolE-Bold9
  <10-12> MnSymbolE-Bold10
  <12->   MnSymbolE-Bold12
}{}

\let\llangle\@undefined
\let\rrangle\@undefined
\DeclareMathDelimiter{\llangle}{\mathopen}%
                     {MnLargeSymbols}{'164}{MnLargeSymbols}{'164}
\DeclareMathDelimiter{\rrangle}{\mathclose}%
                     {MnLargeSymbols}{'171}{MnLargeSymbols}{'171}
\makeatother

\makeatletter
\renewcommand*\email[1][]{\begingroup\sanitize@url\@email{#1}}%
\def\@email#1#2{%
 \endgroup
 \@AF@join{#1#2}%
}%
\makeatother

\begin{document}

\title{Efficient anisotropic Migdal--Eliashberg calculations with the Intermediate Representation basis and Wannier interpolation}

\author{Hitoshi Mori}
\affiliation{Department of Physics, Applied Physics,and Astronomy, Binghamton University-SUNY, Binghamton, New York 13902, USA}
\affiliation{RIKEN Center for Emergent Matter Science, 2-1 Hirosawa, Wako, 351-0198, Japan}
\author{Takuya Nomoto}
\affiliation{Research Center for Advanced Science and Technology, The University of Tokyo, 4-6-1 Komaba, Meguro-ku, Tokyo 153-8904, Japan}
\affiliation{Department of Physics, Tokyo Metropolitan University, Hachioji, Tokyo 192-0397, Japan}
\author{Ryotaro Arita}
\affiliation{RIKEN Center for Emergent Matter Science, 2-1 Hirosawa, Wako, 351-0198, Japan}
\affiliation{Research Center for Advanced Science and Technology, The University of Tokyo, 4-6-1 Komaba, Meguro-ku, Tokyo 153-8904, Japan}
\author{Elena R. Margine}
\affiliation{Department of Physics, Applied Physics,and Astronomy, Binghamton University-SUNY, Binghamton, New York 13902, USA}
\date{\today}
\begin{abstract}

In this study, we combine the \textit{ab initio} Migdal--Eliashberg approach with the intermediate representation for the Green's function, enabling accurate and efficient calculations of the momentum-dependent superconducting gap function while fully considering the effect of the Coulomb retardation. Unlike the conventional scheme that relies on a uniform sampling across Matsubara frequencies - demanding hundreds to thousands of points - the intermediate representation works with fewer than 100 sampled Matsubara Green's functions. The developed methodology is applied to investigate the superconducting properties of three representative low-temperature elemental metals: aluminum (Al), lead (Pb), and niobium (Nb). The results demonstrate the power and reliability of our computational technique to accurately solve the \textit{ab initio} anisotropic Migdal--Eliashberg equations even at extremely low temperatures, below 1~Kelvin.
\end{abstract}

\maketitle

\section{Introduction}

\textit{Ab initio} modeling of conventional superconductors is primarily based on the Eliashberg theory~\cite{Eliashberg1960, Eliashberg1961}, a many-body Green's function approach, where the superconducting pairing arises from an attractive phonon-mediated electron–electron interaction, which overcomes the inter-electron Coulomb repulsion~\cite{Scalapino1966-pr, Scalapino1969book, Allen1983-ko, Carbotte1990-qo}. 
This formalism provides the fundamental equations for calculating the frequency- and temperature-dependent superconducting gap function, but solving the Eliashberg equations from first principles has necessitated several methodological developments~\cite{Marsiglio2008book, Lilia_2022, Pickett2023-fm}. 
For instance, a highly accurate interpolation technique using Wannier functions has enabled the treatment of anisotropy (momentum-dependence) in the electron-phonon interactions~\cite{Margine2013-kc}. 
The Wannier-based algorithm to solve the Eliashberg equations has become an established approach to study the anisotropic nature of the superconducting gap, and has been successfully applied to describe material-specific properties in different families of phonon-mediated superconductors~\cite{Margine2016-nv, Heil2017-hj, Paudyal2020-zx, Dong2022-db, Kafle2022-ji, Tsuppayakorn-Aek2022-wz, Das2023-bp, Lucrezi2024-bm}. 
Methods have also been developed to include the Coulomb interaction from first principles, eliminating the need to reduce it to a single semi-empirical parameter $\mu^*$~\cite{Morel1962-bu, Simonato2023-sg}. 
Notably, recent studies have accounted for electron-phonon and Coulomb interactions on an equal footing when solving the Eliashberg equations within the isotropic approximation~\cite{Davydov2020-xo, Pellegrini2023-vr} or the anisotropic linearized gap equation near the critical temperature~\cite{Wang2020-xw, Wang2021-vh, Wang2023-dx}. 

It remains nevertheless a challenge, within the anisotropic Eliashberg framework, to compute the superconducting gap for low critical temperature ($T_{\mathrm{c}}$) superconductors or to treat both interactions from first principles at the same anisotropic level. One reason behind these limitations is that the Eliashberg equations involve a summation over an extremely large number of Matsubara frequencies as long as a standard representation of the Green's function is used~\cite{Moca2002-it, Sanna2012-rx}. 
In general, the typical energy scale of the electron-phonon interaction is several tens of meV, whereas that of the Coulomb interaction is a few tens of eV. This translates into twenty thousand sampling points for a Matsubara frequency cutoff of approximately 10~eV at $T=1$~K. In addition, since the Coulomb scattering of electrons arising from high-energy states may remain large, contributions from these states also need to be included in the calculations. This requires prohibitively large computational resources as the total numerical cost is determined by the multiplication of the number of electronic states and that of Matsubara frequencies. Therefore, the vast majority of studies have adopted the retarded Coulomb pseudopotential as represented by the Morel--Anderson pseudopotential~\cite{Morel1962-bu} when performing Eliashberg calculations. This approximation effectively reduces the Coulomb interaction to the low-energy scale where the electron-phonon coupling is active and allows us to limit the Matsubara frequency range to less than 1~eV, drastically reducing the numerical cost for the summation over many Matsubara frequencies.

The intermediate representation (IR)~\cite{Shinaoka2017-cg, Li2020-uz, Wallerberger2023-cc} method proposed recently has overcome the long-standing Matsubara frequency sampling problem discussed above. This method is based on the compact representation of the Green's functions both in the imaginary-time and Matsubara-frequency domain, and has been successfully applied to a wide variety of problems~\cite{Nagai2019-oc, Nomoto2020-bp, Nomura2020-tq, Witt2021-gg, Yeh2022-no, Witt2022-aw, Sakurai2022-cl, Nagai2023-yo}, including solving the anisotropic linearized Eliashberg equation in vicinity of the transition temperature~\cite{Wang2020-xw, Wang2021-vh, Wang2023-dx}. The technique allows us to describe the Matsubara Green's function in the frequency range of well over several tens of eV with only around 100 sampling points. 

In the present work, we have combined the IR basis technique with the anisotropic Eliashberg approach utilizing the Wannier-based interpolation to efficiently sample the Green's function and the electron-phonon interaction in the Matsubara frequency domain and the momentum space. This allows describing the superconducting gap functions for high-energy states and high Matsubara frequencies up to several tens of eV at a feasible computational cost. We apply the developed methodology to compute the momentum-dependent superconducting gap function for three representative low-$T_{\mathrm{c}}$ elemental metals (Al, Pb and Nb), and show that the \textit{ab initio} anisotropic Eliashberg equations can be accurately solved at extremely low temperatures, below 1~Kelvin. 

\section{Method}
\subsection{Migdal--Eliashberg equations\label{sec:ME}}

The derivation of the Migdal--Eliashberg (ME) equations using the Nambu--Gor’kov formalism~\cite{Nambu1960-zd, Gorkov1958} and Migdal's theorem~\cite{migdal1958interaction} has been presented in many reviews and textbooks~\cite{Allen1983-ko, PONCE2016116, Davydov2020-xo, Lucrezi2024-bm}. Here, we focus on numerical issues in solving the ME equations. 
The ME equations can be written in terms of the normal self-energy $\Sigma_{n\bm{k}}$ and the pairing self-energy $\phi_{n\bm{k}}$ as: 
\begin{flalign}
&\hspace{0.7em}\Sigma_{n\bm{k}}(\im\omega_{j}^{(\mathrm{F})})
= 
-
\frac{T}{N_{\mathrm{k'}}}
\sum_{\bm{k}',n'}
\sum_{j'}
G_{n'\bm{k}'}(\im\omega_{j'}^{(\mathrm{F})})&
\notag\\
&\hspace{9em}\times
W_{n\bm{k},n'\bm{k}'}(\im\omega_{j}^{(\mathrm{F})}-\im\omega_{j'}^{(\mathrm{F})}),&
\label{Sigma2}
\end{flalign}
\begin{flalign}
&\hspace{0.7em}\phi_{n\bm{k}}(\im\omega_{j}^{(\mathrm{F})})
= 
\frac{T}{N_{\mathrm{k'}}}
\sum_{\bm{k}',n'}
\sum_{j'}
F_{n'\bm{k}'}(\im\omega_{j'}^{(\mathrm{F})})&
\notag\\
&\hspace{9em}\times
W_{n\bm{k},n'\bm{k}'}(\im\omega_{j}^{(\mathrm{F})}-\im\omega_{j'}^{(\mathrm{F})}),&
\label{phi2}
\end{flalign}
\begin{flalign}
&\hspace{0.7em}N_{\e}
=
\frac{1}{N_{\mathrm{k}}}
\sum_{\bm{k},n}
\left[
1+
2T
\sum_{j}
\mathrm{Re}[G_{n\bm{k}}(\im\omega_{j'}^{(\mathrm{F})})]
\right],&
\label{ne2}
\end{flalign}
where $T$ is the temperature, $W_{n\bm{k},n'\bm{k}'}$ is a pairing interaction, $\im\omega_{j}^{(\mathrm{F})}=(2n_j + 1)\im\pi T$ stands for the fermionic Matsubara frequency with $n_j$ being an integer, and $N_{\mathrm{k}}$ and $N_{\mathrm{k'}}$ are the number of wavevectors $\bm{k}$ and $\bm{k}'$ in the first Brillouin zone. 
$G_{n'\bm{k}'}(\im\omega_{j'}^{(\mathrm{F})})$ and $F_{n'\bm{k}'}(\im\omega_{j'}^{(\mathrm{F})})$ represent the normal and anomalous Green's functions for electrons. Equation~\eqref{ne2} serves as a constraint for the two self-energy equations to ensure the conservation of the electron number per unit cell, denoted as $N_{\e}$. In the following, we choose the gauge such that the anomalous Green's function is a real-valued function. 

The kernel $W_{n\bm{k},n'\bm{k}'}$ consists of the contributions from the interaction due to the electron-phonon coupling and the screened Coulomb interaction as
\begin{align}
W_{n\bm{k},n'\bm{k}'} = W_{n\bm{k},n'\bm{k}'}^{\text{el-ph}} + W_{n\bm{k},n'\bm{k}'}^{\mathrm{C}}.
\end{align}
In the Migdal's approximation~\cite{migdal1958interaction}, the electron-phonon interaction $W_{n\bm{k},n'\bm{k}'}^{\text{el-ph}}$ can be written as: 
\begin{align}
W_{n\bm{k},n'\bm{k}'}^{\text{el-ph}}(\im\omega_{j}^{(\mathrm{B})})
=
\sum_{\nu}
|g_{n\bm{k},n'{\bm{k}'}}^{\nu}|^2
D_{\nu\bm{k}-\bm{k}'}(\im\omega_{j}^{(\mathrm{B})}),
\label{el-ph-int}
\end{align}
where $g_{n\bm{k},n'{\bm{k}'}}^{\nu}$ is the electron-phonon matrix element, $D_{\nu\bm{q}}$ is the dressed phonon Green's function, and $\im\omega_{j}^{(\mathrm{B})}=2n_j\im\pi T$ is the bosonic Matsubara frequency. 
Due to momentum conservation, $\bm{q} = \bm{k}-\bm{k}'$ is a phonon wavevector.
In this work, we ignore renormalization effects of the phonon propagator and replace $D_{\nu\bm{q}}$ with the bare phonon Green's function, 
\begin{align}
&D^{(0)}_{\nu\bm{q}}(\im\omega_{j}^{(\mathrm{B})})
=
-\frac{
2\omega_{\nu\bm{q}}
}{
[\omega_{j}^{(\mathrm{B})}]^2+\omega_{\nu\bm{q}}^2
},\label{phgreen}
\end{align}
where $\omega_{\nu\bm{q}}$ is the phonon frequency. 
Here, we introduce the even functions of $\im\omega_{j'}^{(\mathrm{F})}$, namely the mass renormalization factor $Z_{n\bm{k}}(\im\omega_{j}^{(\mathrm{F})})$ and the energy shift $\chi_{n\bm{k}}(\im\omega_{j}^{(\mathrm{F})})$. 
These are defined as linear combinations of the normal self-energy, $\Sigma_{n\bm{k}}(\im\omega_{j}^{(\mathrm{F})})$: 
\begin{align}
\im\omega_{j}^{(\mathrm{F})}[1-Z_{n\bm{k}}(\im\omega_{j}^{(\mathrm{F})})]
&= 
\frac{1}{2}
[\Sigma_{n\bm{k}}(\im\omega_{j}^{(\mathrm{F})})-\Sigma_{n\bm{k}}(-\im\omega_{j}^{(\mathrm{F})})], 
\label{znorm-sigma}\\
\chi_{n\bm{k}}(\im\omega_{j}^{(\mathrm{F})})
&= 
\frac{1}{2}
[\Sigma_{n\bm{k}}(\im\omega_{j}^{(\mathrm{F})})+\Sigma_{n\bm{k}}(-\im\omega_{j}^{(\mathrm{F})})].
\label{chi-sigma}
\end{align}
Using $Z_{n\bm{k}}(\im\omega_{j}^{(\mathrm{F})})$, $\chi_{n\bm{k}}(\im\omega_{j}^{(\mathrm{F})})$, and $\phi_{n\bm{k}}(\im\omega_{j}^{(\mathrm{F})})$, we can express the normal and anomalous Green's functions. 
\begin{align}
&G_{n\bm{k}}(\im\omega_{j}^{(\mathrm{F})})
=
-\frac{
\im\omega_{j}^{(\mathrm{F})}Z_{n\bm{k}}(\im\omega_{j}^{(\mathrm{F})}) + \varepsilon_{n\bm{k}}-\varepsilon_{\text{F}}+\chi_{n\bm{k}}(\im\omega_{j}^{(\mathrm{F})})
}{
\Theta_{n\bm{k}}(\im\omega_{j}^{(\mathrm{F})})
},\label{normal}\\
&F_{n\bm{k}}(\im\omega_{j}^{(\mathrm{F})})
=
-\frac{
\phi_{n\bm{k}}(\im\omega_{j}^{(\mathrm{F})})
}{
\Theta_{n\bm{k}}(\im\omega_{j}^{(\mathrm{F})})
}\label{anomalous}
\end{align}
with
\begin{flalign}
&\hspace{0.7em}\Theta_{n\bm{k}}(\im\omega_{j}^{(\mathrm{F})})
\notag\\
&\hspace{1em}
=[\omega_{j}^{(\mathrm{F})}Z_{n\bm{k}}(\im\omega_{j}^{(\mathrm{F})})]^2
+[\varepsilon_{n\bm{k}}-\varepsilon_{\text{F}}+\chi_{n\bm{k}}(\im\omega_{j}^{(\mathrm{F})})]^2&
\notag\\
&\hspace{9em}
+[\phi_{n\bm{k}}(\im\omega_{j}^{(\mathrm{F})})]^2,&
\end{flalign}
where $\varepsilon_{n\bm{k}}$ is the electronic eigenenergy. 
When solving Eqs.~\eqref{Sigma2} and \eqref{phi2} for $\Sigma_{n\bm{k}}$ and $\phi_{n\bm{k}}$ self-consistently, the Fermi level $\varepsilon_{\text{F}}$ is updated to ensure the charge neutrality of the system, Eq.~\eqref{ne2}. 
Applying Eqs.~\eqref{znorm-sigma}--\eqref{anomalous} into Eqs.~\eqref{Sigma2}--\eqref{ne2} yields another expression of the ME equations: 
\begin{flalign}
&\hspace{0.7em}Z_{n\bm{k}}(\im\omega_{j}^{(\mathrm{F})})&
\notag\\
&\hspace{1em}
= 
1
-\frac{T}{\omega_{j}^{(\mathrm{F})} N_{\mathrm{k'}}}
\sum_{\bm{k}',n'}
\sum_{j'}
\frac{\omega_{j'}^{(\mathrm{F})}Z_{n'\bm{k}'}(\im\omega_{j'}^{(\mathrm{F})})}{\Theta_{n'\bm{k}'}(\im\omega_{j'}^{(\mathrm{F})})}&
\notag\\
&\hspace{9em}\times
W_{n\bm{k},n'\bm{k}'}(\im\omega_{j}^{(\mathrm{F})}-\im\omega_{j'}^{(\mathrm{F})}),&
\label{znorm1}
\end{flalign}
\begin{flalign}
&\hspace{0.7em}\chi_{n\bm{k}}(\im\omega_{j}^{(\mathrm{F})})&
\notag\\
&\hspace{1em}
= 
\frac{T}{N_{\mathrm{k'}}}
\sum_{\bm{k}',n'}
\sum_{j'}
\frac{\varepsilon_{n'\bm{k}'}-\varepsilon_{\text{F}}+\chi_{n'\bm{k}'}(\im\omega_{j'}^{(\mathrm{F})})}{\Theta_{n'\bm{k}'}(\im\omega_{j'}^{(\mathrm{F})})}&
\notag\\
&\hspace{9em}\times
W_{n\bm{k},n'\bm{k}'}(\im\omega_{j}^{(\mathrm{F})}-\im\omega_{j'}^{(\mathrm{F})}),&
\label{chi1}
\end{flalign}
\begin{flalign}
&\hspace{0.7em}\phi_{n\bm{k}}(\im\omega_{j}^{(\mathrm{F})})&
\notag\\
&\hspace{1em}
= 
-\frac{T}{N_{\mathrm{k'}}}
\sum_{\bm{k}',n'}
\sum_{j'}
\frac{\phi_{n'\bm{k}'}(\im\omega_{j'}^{(\mathrm{F})}) }{\Theta_{n'\bm{k}'}(\im\omega_{j'}^{(\mathrm{F})})}
W_{n\bm{k},n'\bm{k}'}(\im\omega_{j}^{(\mathrm{F})}-\im\omega_{j'}^{(\mathrm{F})}),&
\label{phi1}
\\
&\hspace{0.7em}N_{\e}
=
\frac{1}{N_{\mathrm{k}}}
\sum_{\bm{k},n}
\left[
1-
2T
\sum_{j}
\frac{\varepsilon_{n\bm{k}}-\varepsilon_{\text{F}}+\chi_{n\bm{k}}(\im\omega_{j}^{(\mathrm{F})}) }{\Theta_{n\bm{k}}(\im\omega_{j}^{(\mathrm{F})})}
\right].&
\label{ne1}
\end{flalign}
Equations~\eqref{znorm1}--\eqref{ne1} are essentially equivalent to Eqs.~\eqref{Sigma2}--\eqref{ne2}. 
The superconducting gap function $\Delta_{n\bm{k}}(\im\omega_{j}^{(\mathrm{F})})$ is determined using $\phi_{n\bm{k}}(\im\omega_{j}^{(\mathrm{F})})$ as 
\begin{align}
\Delta_{n\bm{k}}(\im\omega_{j}^{(\mathrm{F})})=
\frac{
\phi_{n\bm{k}}(\im\omega_{j}^{(\mathrm{F})})
}{
Z_{n\bm{k}}(\im\omega_{j}^{(\mathrm{F})})
}.
\end{align}
\par
In this work, we introduce a dimensionless Coulomb interaction parameter $\mu_{\mathrm{C}}$ defined as the double Fermi-surface average over $\bm{k}$ and $\bm{k}'$ of the
Coulomb interaction $W^{\mathrm{C}}$: 
\begin{align}
\mu_{\mathrm{C}} 
= 
N_{\mathrm{F}}
\left\llangle
W_{n\bm{k},n'\bm{k}'}^{\mathrm{C}}(\im \omega_{j}=0)
\right\rrangle_{\text{F.S.}},
\end{align}
where $N_{\mathrm{F}}$ represents the density of states per spin at the Fermi energy, $\varepsilon_{\text{F}}^{(0)}$, which is calculated using the Gaussian smearing method for the non-interacting state at zero temperature. 
We also apply the approximation $W\simeq W^{\text{el-ph}}$ to Eq.~\eqref{Sigma2}. 
This can be justified by the following two facts: 
(1) Terms operated by the static Coulomb interaction ($\mu_{\mathrm{C}}/N_{\mathrm{F}}$) have no contribution to $Z_{n\bm{k}}$, and (2) the Coulomb correction to the energy shift $\chi_{n\bm{k}}$ is expected to be small since the eigenenergy $\varepsilon_{n\bm{k}}$ obtained from first-principles calculations partially includes some contribution from Coulomb interactions~\cite{Allen1983-ko}. 
To account for the Coulomb interactions from high-energy states, we introduce two different energy windows and divide the summation over the final states $(\bm{k}',n')$ in Eq.~\eqref{phi2}. 
Finally, the equations we solve in this work are as follows: 
\begin{flalign}
&\hspace{0.7em}\Sigma_{n\bm{k}}(\im\omega_{j}^{(\mathrm{F})})&
\notag\\
&\hspace{1em}
= 
-
\frac{T}{N_{\mathrm{q}}}
\sum_{\bm{q},n'}^{\mathrm{Inner.}}
\sum_{j'}
G_{n'\bm{k}+\bm{q}}(\im\omega_{j'}^{(\mathrm{F})})
W^{\text{el-ph}}_{n\bm{k},n'\bm{k}+\bm{q}}(\im\omega_{j}^{(\mathrm{F})}-\im\omega_{j'}^{(\mathrm{F})}),&
\label{Sigma3}
\end{flalign}
\begin{flalign}
&\hspace{0.7em}\phi_{n\bm{k}}(\im\omega_{j}^{(\mathrm{F})})&
\notag\\
&\hspace{1em}
= 
\frac{T}{N_{\mathrm{q}}}
\sum_{\bm{q},n'}^{\mathrm{Inner.}}
\sum_{j'}
F_{n'\bm{k}+\bm{q}}(\im\omega_{j'}^{(\mathrm{F})})
W^{\text{el-ph}}_{n\bm{k},n'\bm{k}+\bm{q}}(\im\omega_{j}^{(\mathrm{F})}-\im\omega_{j'}^{(\mathrm{F})})&
\notag\\
&\hspace{3em}
+\frac{T}{N_{\mathrm{k}_\mathrm{C}}}
\Biggl[
\sum_{\bm{k}_{\mathrm{C}},n'}^{\mathrm{Inner.}}
\sum_{j'}
F_{n'\bm{k}_{\mathrm{C}}}(\im\omega_{j'}^{(\mathrm{F})})
\frac{\mu_{\mathrm{C}}}{N_{\mathrm{F}}}
\Biggr.&
\notag\\
&\hspace{9em}
\Biggl.
+
\sum_{\bm{k}_{\mathrm{C}},n'}^{\mathrm{Outer.}}
\sum_{j'}
F^{\mathrm{(out)}}_{n'\bm{k}_{\mathrm{C}}}(\im\omega_{j'}^{(\mathrm{F})})
\frac{\mu_{\mathrm{C}}}{N_{\mathrm{F}}}
\Biggr],&
\label{phi3}
\end{flalign}
\begin{flalign}
&\hspace{0.7em}\phi^{\mathrm{(out)}}_{n\bm{k}}(\im\omega_{j}^{(\mathrm{F})})&
\notag\\
&\hspace{1em}
= 
\frac{T}{N_{\mathrm{k}_\mathrm{C}}}
\Biggl[
\sum_{\bm{k}_{\mathrm{C}},n'}^{\mathrm{Inner.}}
\sum_{j'}
F_{n'\bm{k}_{\mathrm{C}}}(\im\omega_{j'}^{(\mathrm{F})})
\frac{\mu_{\mathrm{C}}}{N_{\mathrm{F}}}
\Biggr.&
\notag\\
&\hspace{7em}
\Biggl.
+
\sum_{\bm{k}_{\mathrm{C}},n'}^{\mathrm{Outer.}}
\sum_{j'}
F^{\mathrm{(out)}}_{n'\bm{k}_{\mathrm{C}}}(\im\omega_{j'}^{(\mathrm{F})})
\frac{\mu_{\mathrm{C}}}{N_{\mathrm{F}}}
\Biggr],&
\label{tphi3}
\end{flalign}
with
\begin{flalign}
&\hspace{0.7em}F^{\mathrm{(out)}}_{n\bm{k}}(\im\omega_{j}^{(\mathrm{F})})
=
-\frac{
\phi^{\mathrm{(out)}}_{n\bm{k}}(\im\omega_{j}^{(\mathrm{F})})
}{
\Theta^{\mathrm{(out)}}_{n\bm{k}}(\im\omega_{j'}^{(\mathrm{F})})
}&\label{out-anomalous}
\end{flalign}
and
\begin{flalign}
&\hspace{0.7em}\Theta^{\mathrm{(out)}}_{n\bm{k}}(\im\omega_{j}^{(\mathrm{F})})&
\notag\\
&\hspace{1em}
=[\omega_{j}^{(\mathrm{F})}]^2
+[\varepsilon_{n\bm{k}}-\varepsilon_{\text{F}}]^2
+[\phi^{\mathrm{(out)}}_{n\bm{k}}(\im\omega_{j}^{(\mathrm{F})})]^2,&
\end{flalign}
where $\bm{q}$ is the scattering wavevector for the electron-phonon interaction, while $\bm{k}_{\mathrm{C}}$ is a newly introduced wavevector for the Coulomb interaction. 
$N_{\mathrm{q}}$ and $N_{\mathrm{k}_\mathrm{C}}$ are the number of wavevectors $\bm{q}$ and $\bm{k}_{\mathrm{C}}$ in the first Brillouin zone. 
The $\bm{k}_{\mathrm{C}}$ grid is required to be commensurate with and smaller than (or equal to) the $\bm{k}$ grid in order to map the two grids during the self-consistent procedure in the ME equations. 
$F_{n\bm{k}}^{\mathrm{(out)}}$ and $\phi_{n\bm{k}}^{\mathrm{(out)}}$ are the anomalous Green's function and the pairing self-energy defined outside the inner window on the $\bm{k}_{\mathrm{C}}$ grid~\footnote{The eigenenergies for the states outside the inner window required for calculating $F_{n\bm{k}}^{\mathrm{(out)}}$ are obtained from a non-self-consistent DFT calculation on the $\bm{k}_{\mathrm{C}}$ grid.}. 
$\sum_{\bm{q},n'}^{\mathrm{Inner.}}$ represents the summation over the final states $(\bm{k}+\bm{q}, n')$ lying within the inner window. 
Similarly, $\sum_{\bm{k}_{\mathrm{C}},n'}^{\mathrm{Inner.}}$ and $\sum_{\bm{k}_{\mathrm{C}},n'}^{\mathrm{Outer.}}$ denote the summations over the final states $(\bm{k}_{\mathrm{C}}, n')$ lying within and outside the inner window, respectively. 
Outside the inner window, we approximate $Z_{n\bm{k}}(\im\omega_{j}^{(\mathrm{F})})\sim 1$, $\chi_{n\bm{k}}(\im\omega_{j}^{(\mathrm{F})})\sim 0$, and $F_{n'\bm{k}+\bm{q}}(\im\omega_{j'}^{(\mathrm{F})})
W^{\text{el-ph}}_{n\bm{k},n'\bm{k}+\bm{q}}(\im\omega_{j}^{(\mathrm{F})}-\im\omega_{j'}^{(\mathrm{F})})\sim 0$. 
At this point it is important to emphasize that the Coulomb parameter $\mu_{\mathrm{C}}$ appearing in Eqs.~\eqref{phi3} and \eqref{tphi3} is not a ``retarded'' Coulomb pseudopotential, which is typified by the Morel--Anderson pseudopotential~\cite{Morel1962-bu}. 
By enlarging the outer window, these equations can take into account the Coulomb retardation effect coming from higher energy states, while limiting the energy region where we calculate the electron-phonon interaction to the inner window. 
In Eq.~\eqref{ne2}, $\sum_{\bm{k},n}$ is also replaced by $\sum_{\bm{k},n}^{\mathrm{Inner.}}$, which represents the summation over the states within the inner window. \par

The cutoff in the Matsubara summation is generally determined by the frequency dependence of $W$. 
Particularly, in phonon-driven superconductors, the typical cutoff is a few times the maximum phonon frequency because the electron-phonon interaction mainly contributes to $W$, and can be regarded as a linear combination of Lorentzian functions with the phonon frequency $\omega_{\nu\bm{q}}$ being the half-width at half-maximum. 
In addition, employing a conventional uniform grid for Matsubara frequencies leads to an increase in the required number of Matsubara points at a rate of $T^{-1}$ as the temperature decreases, resulting in extremely high computational costs. 
Consequently, calculating the superconducting gap functions for systems with transition temperatures below 10~K has posed significant challenges in terms of both computational and memory requirements. Previous studies have managed to cut down computational costs to some extent by employing the fast Fourier transform (FFT) algorithm~\cite{Monthoux1993-vy,Wermbter1993-ok,Pao1995-gw,Dolgov2005-qu}. 
However, the FFT algorithm does not address the problem of memory shortage. For calculations, especially at temperatures below 1~K, we need to introduce innovative methods, such as the IR method described below.

\subsection{Fourier transformation with the IR basis}

To solve Eqs.~\eqref{Sigma2}--\eqref{ne2}, we have to calculate the convolution of $G$ and $W$ and that of $F$ and $W$ for all sampled Matsubara frequencies. 
Using the convolution property of the Fourier transformation, the convolution of two functions $f$ and $g$ can be written as 
\begin{align}
\sum_{j}f(\im\omega_{j}-\im\omega_{j'})g(\im\omega_{j'})=\mathcal{F}^{-1}[\mathcal{F}(f)\cdot\mathcal{F}(g)],
\label{convl}
\end{align}
where $\mathcal{F}$ and $\mathcal{F}^{-1}$ denote the Fourier and inverse Fourier transformations between the Matsubara frequency domain and the imaginary time domain. 
To reduce the cost for computing the convolutions on the Matsubara frequency, we introduce the IR basis functions. In practical calculations, we compute sampling points of imaginary time and Matsubara frequency, $\{\bar{\tau}_i\}$, $\{\bar{\omega}_{j}^{(\mathrm{F})}\}$, and $\{\bar{\omega}_{j}^{(\mathrm{B})}\}$, and the IR basis functions, $U_l(\bar{\tau}_i)$, $\hat{U}_l^{(\mathrm{F})}(\im\bar{\omega}_j^{(\mathrm{F})})$, and $\hat{U}_l^{(\mathrm{B})}(\im\bar{\omega}_j^{(\mathrm{B})})$ prior to solving the ME equations. 
Using them, one can obtain the following compact representation of the Green's function: 
\begin{align}
G(\im\bar{\omega}_j^{(\mathrm{F})})&
=
\sum_{l}
G_l\hat{U}_l^{(\mathrm{F})}(\im\bar{\omega}_j^{(\mathrm{F})}),
\label{IR-G1}
\\
G(\bar{\tau}_i)&
=
\sum_{l}
G_lU_l(\bar{\tau}_i),
\label{IR-G2}
\end{align}
where $G_l$ are the IR expansion coefficients. 
We can evaluate $G_l$ from the functions $G(\im\bar{\omega}_j^{(\mathrm{F})})$ and $G(\bar{\tau}_i)$ by using the pseudo-inverse matrix, $\mathbf{U}^{+}\equiv (\mathbf{U}^\dagger\mathbf{U})^{-1} \mathbf{U}^\dagger$: 
\begin{align}
G_l& =
\sum_{j}
\left[
\hat{\mathbf{U}}_{(\mathrm{F})}^{+}
\right]_{lj}
G(\im\bar{\omega}_j^{(\mathrm{F})})
\label{IR-G3}
\\
& =
\sum_{i}
\left[
\mathbf{U}^{+}
\right]_{li}
G(\bar{\tau}_i),
\label{IR-G4}
\end{align}
where $[\hat{\mathbf{U}}_{(\mathrm{F})}]_{jl}=\hat{U}_l^{(\mathrm{F})}(\im\bar{\omega}_j^{(\mathrm{F})})$ and $[\mathbf{U}]_{il}=U_l(\bar{\tau}_i)$. 
Since $W$ is a bosonic function, we can consider similar relations to Eqs.~\eqref{IR-G1}--\eqref{IR-G4} by replacing $\hat{U}_l^{(\mathrm{F})}(\im\bar{\omega}_j^{(\mathrm{F})})$ with $\hat{U}_l^{(\mathrm{B})}(\im\bar{\omega}_j^{(\mathrm{B})})$. \par
The sampling points and the IR basis functions can be computed using the \textsc{sparse-ir} library~\cite{Shinaoka2017-cg, Li2020-uz, Wallerberger2023-cc}. 
Two parameters, $\Lambda$ and $\epsilon_{\mathrm{IR}}$, are used to generate the sampling points. $\Lambda=\omega_{\mathrm{max}}/T$ determines the real frequency domain $[-\omega_{\mathrm{max}};\omega_{\mathrm{max}}]$ of the kernel $K(\tau, \omega)$ in the spectral representation of the Green's function,
\begin{align}
G(\tau)
=
-
\int_{-\omega_{\mathrm{max}}}^{\omega_{\mathrm{max}}}d\omega
K(\tau, \omega)\rho(\omega), 
\label{spcrep}
\end{align}
where $\rho(\omega)$ is the spectral function. 
The IR basis functions are computed by a singular value expansion of the kernel $K(\tau, \omega)$. 
A sufficiently large value of $\Lambda$ must be chosen according to the typical energy scale of the system and the minimum temperature so that the domain $[-\omega_{\mathrm{max}};\omega_{\mathrm{max}}]$ includes all real frequencies where $\rho(\omega)$ has a nonzero value.
$\epsilon_{\mathrm{IR}}$ controls the sparseness of the sampling points, i.e., the numerical accuracy of the Green's function. 

\begin{figure}[!b]
  \begin{center}
    \includegraphics[width=6.6cm]{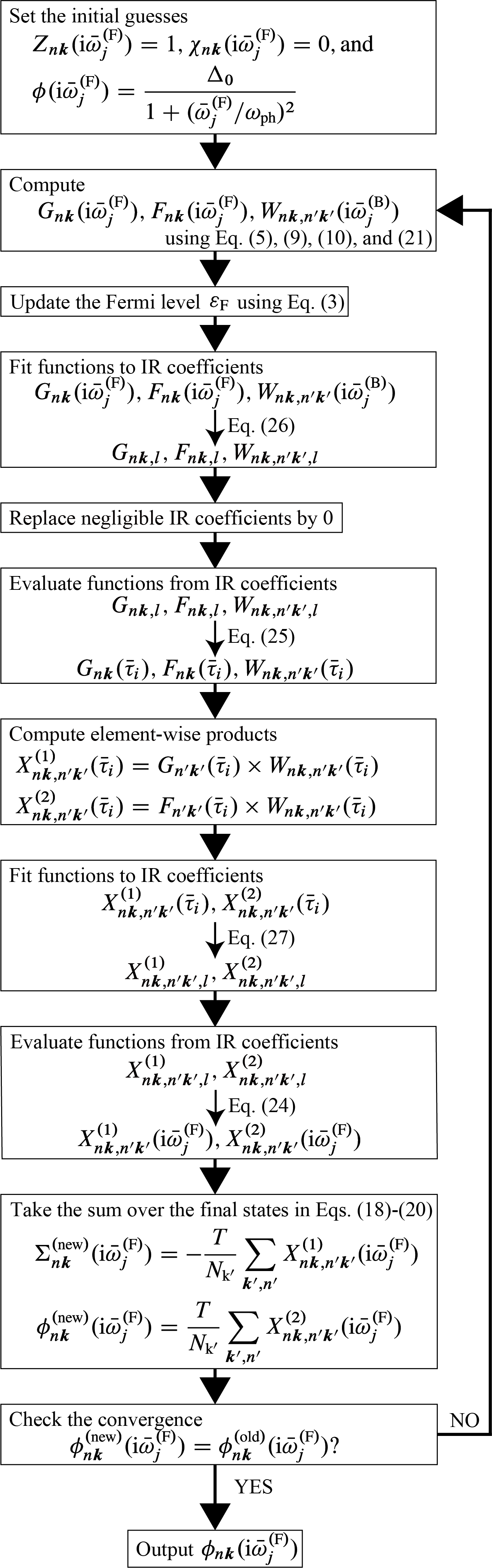}
    \caption{\label{fig:1}The flow chart to solve the ME equations self-consistently using the IR basis.}
    \label{flow}
 \end{center}
\end{figure}

\subsection{Calculation flow\label{sec:calc_details}}

\begin{figure*}
  \begin{center}
    \includegraphics[width=16cm]{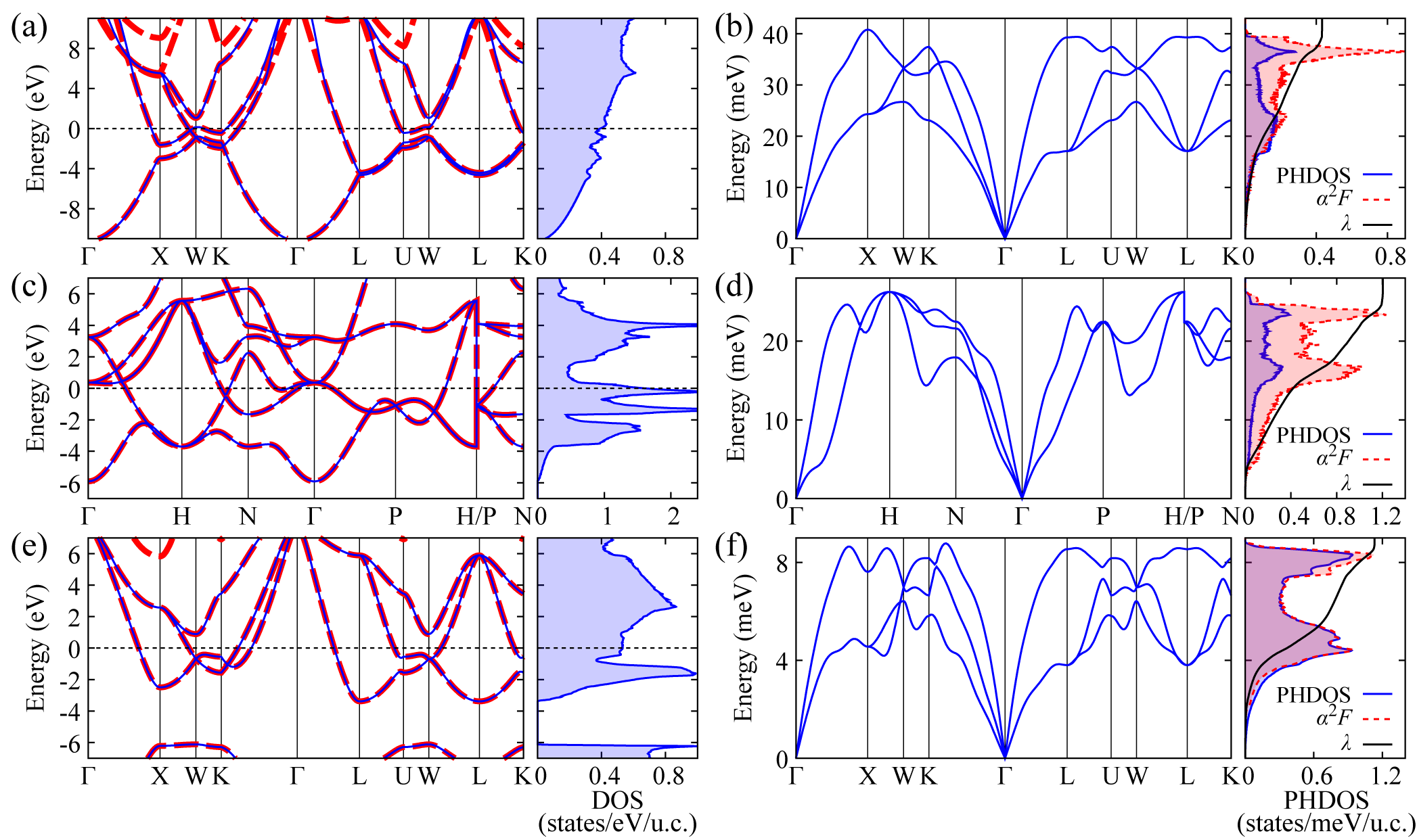}
    \caption{\label{fig:2}Panel~(a) shows the calculated electronic band structure and the density of states (DOS) with respect to the Fermi energy for Al. The dashed red lines represent the DFT bands, and the solid blue lines represent the Wannier bands. Panel~(b) shows the phonon dispersion and the phonon density of states (PHDOS), the isotropic Eliashberg spectral function $\alpha^2 F(\omega)$, and the cumulative electron-phonon coupling strength $\lambda(\omega)$. Panels (c)--(d) and (e)--(f) show the corresponding results for Nb and Pb, respectively.}
 \end{center}
\end{figure*}
\begin{figure*}
  \begin{center}
    \includegraphics[width=16cm]{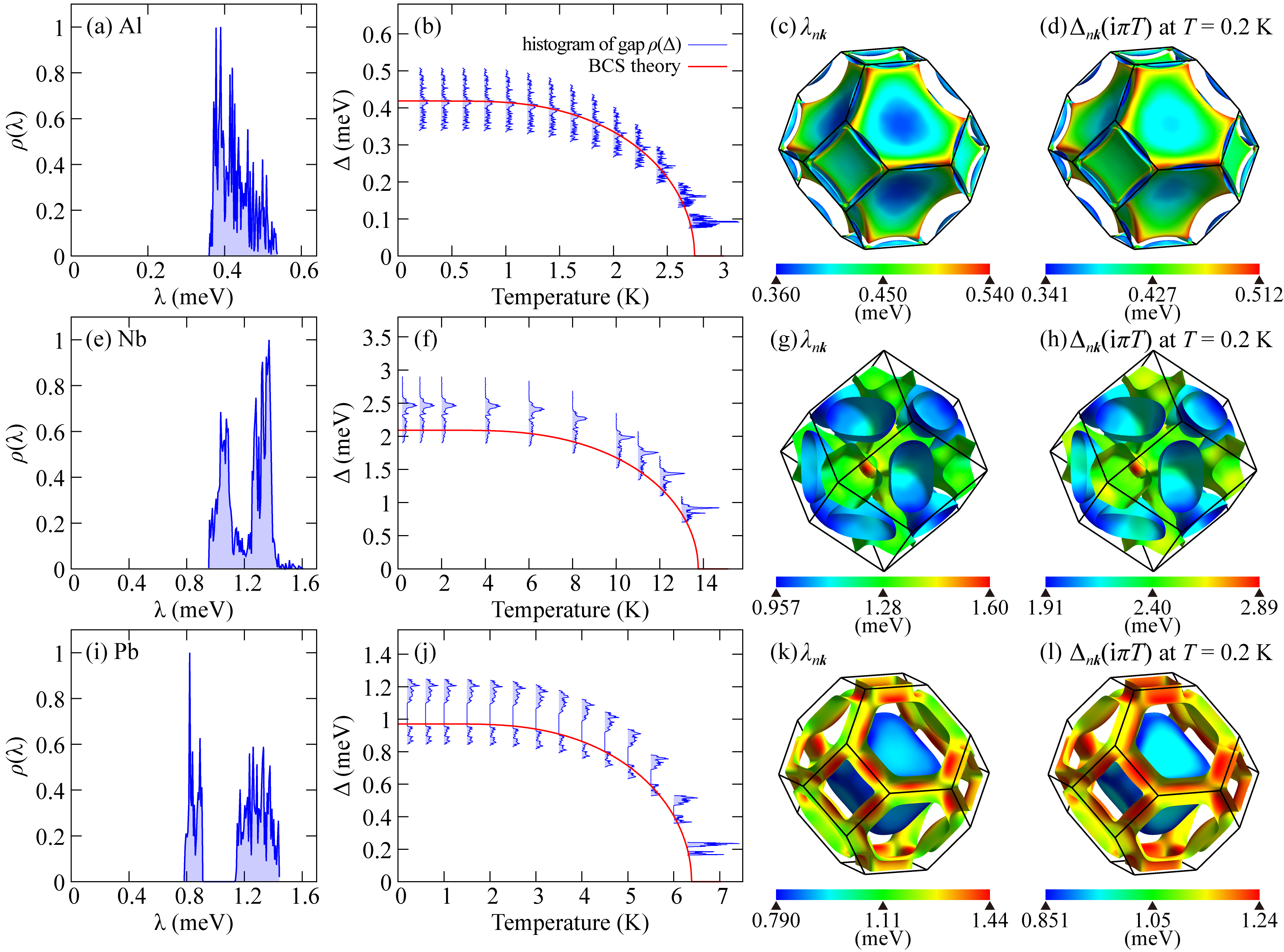}
    \caption{\label{fig:3}Panel~(a) shows the histogram of the state-dependent electron-phonon coupling strength $\rho(\lambda)$. Panel~(b) shows the histograms of the superconducting gap function $\rho(\Delta)$ at the lowest Matsubara frequency for different temperatures. Solid red line represents the temperature dependence of the superconducting gap expected from the BCS theory in the weak coupling limit. Panels~(c)--(d) show the state-dependent electron-phonon coupling strength and the superconducting gap function on the Fermi surface. Panels (e)--(h) and (i)--(l) show the corresponding results for Nb and Pb, respectively. The calculations were performed with $96^3$ $\bm{k}$ and $\bm{q}$ grids, a $48^3$ $\bm{k}_{\rm C}$ grid, and an inner window of 0.5~eV. The images on the Fermi surface were rendered using the \textsc{FermiSurfer} software~\cite{Kawamura2019-ym}.}
 \end{center}
\end{figure*}

Figure~\ref{fig:1} shows the calculation flow chart for solving the anisotropic ME equations. 
The implementation is based on the assumption that the symmetry relation $G(\im\omega_{j}^{(\mathrm{F})})=G^*(-\im\omega_{j}^{(\mathrm{F})})$ always holds so that only summations over the positive Matsubara frequencies need to be considered. 
The ME calculations for each temperature are performed as follows: \\
\indent(1) Set the initial guess: $Z_{n\bm{k}}(\im\bar{\omega}_j^{(\mathrm{F})})=1$, $\chi_{n\bm{k}}(\im\bar{\omega}_j^{(\mathrm{F})})=0$, and 
\begin{align}
\phi_{n\bm{k}}(\im\bar{\omega}_j^{(\mathrm{F})}) 
= 
\frac{\Delta_0}{1+(\bar{\omega}_j^{(\mathrm{F})}/\omega_{\mathrm{ph}})^2},
\end{align}
where $\omega_{\mathrm{ph}}=1.1 \,\times$ (the maximum phonon frequency), and $\Delta_0$ should be set to an appropriate value. 
For a sufficiently low temperature, we set $\Delta_0=1.76\times T_{\mathrm{c}}^{(\mathrm{AD})}$, with $T_{\mathrm{c}}^{(\mathrm{AD})}$ being the Allen--Dynes estimate of the transition temperature~\cite{Allen1975-ad}. \\
\indent(2) Compute the Green's functions $G_{n\bm{k}}(\im\bar{\omega}_j^{(\mathrm{F})})$, $F_{n\bm{k}}(\im\bar{\omega}_j^{(\mathrm{F})})$ and $F_{n\bm{k}}^{\mathrm{(out)}}(\im\bar{\omega}_j^{(\mathrm{F})})$ and the interaction term $W_{n\bm{k},n'\bm{k}'}(\im\bar{\omega}_j^{(\mathrm{B})})$ on the discrete imaginary Mastubara frequency grid based on Eqs.~\eqref{el-ph-int}, \eqref{normal}, \eqref{anomalous}, and \eqref{out-anomalous} using $Z_{n\bm{k}}(\im\bar{\omega}_j^{(\mathrm{F})})$, $\chi_{n\bm{k}}(\im\bar{\omega}_j^{(\mathrm{F})})$, $\phi_{n\bm{k}}(\im\bar{\omega}_j^{(\mathrm{F})})$, and $\phi_{n\bm{k}}^{\mathrm{(out)}}(\im\bar{\omega}_j^{(\mathrm{F})})$.\\
\indent(3) Compute the Fermi level $\varepsilon_{\text{F}}$ such that Eq.~\eqref{ne2} is satisfied using Brent's method~\cite{brent1973algorithms}. \\
\indent(4) Evaluate the IR coefficients $G_{n\bm{k},l}$, $F_{n\bm{k},l}$, $F_{n\bm{k},l}^{\mathrm{(out)}}$, and $W_{n\bm{k},n'\bm{k}',l}$ using Eq.~\eqref{IR-G3}. \\
\indent(5) Replace negligibly small IR coefficients with 0. 
This is equivalent to a noise reduction to prevent truncation errors from accumulating in the iterative procedure. 
If $|G_{n\bm{k},l}|<\epsilon_{\mathrm{cut}}\times \underset{l}{\mathrm{max}}|G_{n\bm{k},l}|$, the value is replaced with 0. A value of $\epsilon_{\mathrm{cut}} = 10^{-5}$ is employed in this work. The same treatment is applied to $F_{n\bm{k},l}$, $F_{n\bm{k},l}^{\mathrm{(out)}}$, and $W_{n\bm{k},n'\bm{k}',l}$. \\
\indent(6) Evaluate the Green's functions $G_{n\bm{k}}(\bar{\tau}_i)$ and $F_{n\bm{k}}(\bar{\tau}_i)$, and the interaction term $W_{n\bm{k},n'\bm{k}'}(\bar{\tau}_i)$ in the imaginary time domain from the IR coefficients using Eq.~\eqref{IR-G2}.\\
\indent(7) Compute the element-wise products 
\begin{align}
X^{(1)}_{n\bm{k},n'\bm{k}'}(\bar{\tau}_i)&
=
G_{n\bm{k}}(\bar{\tau}_i)W_{n\bm{k},n'\bm{k}'}(\bar{\tau}_i), \\
X^{(2)}_{n\bm{k},n'\bm{k}'}(\bar{\tau}_i)&
=
F_{n\bm{k}}(\bar{\tau}_i)W_{n\bm{k},n'\bm{k}'}(\bar{\tau}_i)
\label{convol}
\end{align}
in the imaginary time domain. \\
\indent(8) Extract the IR coefficients of $X^{(1)}_{n\bm{k},n'\bm{k}', l}$ and $X^{(2)}_{n\bm{k},n'\bm{k}', l}$ using Eq.~\eqref{IR-G4}. \\
\indent(9) Evaluate $X^{(1)}_{n\bm{k},n'\bm{k}'}(\im\bar{\omega}_j^{(\mathrm{F})})$ and $X^{(2)}_{n\bm{k},n'\bm{k}'}(\im\bar{\omega}_j^{(\mathrm{F})})$ in the Matsubara frequency space using Eq.~\eqref{IR-G1}. \\
\indent(10) Compute the self-energies $\Sigma_{n\bm{k}}(\im\bar{\omega}_{j}^{(\mathrm{F})})$, $\phi_{n\bm{k}}(\im\bar{\omega}_{j}^{(\mathrm{F})})$, and $\phi_{n\bm{k}}^{\mathrm{(out)}}(\im\bar{\omega}_{j}^{(\mathrm{F})})$ in the Matsubara frequency domain by taking the summation over the final states using Eqs.~\eqref{Sigma3}--\eqref{tphi3}. \\
\indent(11) Check whether the difference between $\phi(\im\bar{\omega}_{j}^{(\mathrm{F})})$ at the current and previous iteration is below a given convergence threshold. Repeat steps (2)--(11) until $\phi(\im\bar{\omega}_{j}^{(\mathrm{F})})$ converges.\par
Here, the interaction term is denoted simply as $W_{n\bm{k},n'\bm{k}'}(\im\bar{\omega}_{j}^{(\mathrm{B})})$ for general notation. However, the above description applies specifically to the convolution with the electron-phonon interaction since the Coulomb interaction is treated in the static approximation as $W^{\mathrm{C}}=\mu_{\mathrm{C}}/N_{\mathrm{F}}$. 
See also Appendix~\ref{sec:matsu} for a discussion of the treatment of the static term. 

\section{Results and discussion}

In this paper, we calculate the superconducting gap function of three elemental materials, fcc Al, bcc Nb, and fcc Pb. 
All the first-principles calculations are performed with the \textsc{Quantum espresso} package~\cite{QE-2009,QE-2017,QE-2020}. 
We use the Perdew-Burke-Ernzerhof parametrization of the generalized gradient approximation (PBE-GGA)~\cite{Perdew1996-mv} and the optimized norm-conserving Vanderbilt (ONCV) pseudopotentials~\cite{Hamann2013-cq} extracted from the SG15 ONCV library~\cite{Schlipf2015-xb}. The plane wave cutoff is set to 100~Ry. 
The optimized lattice parameters are found to be $a=$ 4.04, 3.31 and 5.03~$\mathring{\mathrm{A}}$ for Al, Nb, and Pb, respectively. 
The dynamical matrices and the linear variation of the self-consistent potential are calculated within density-functional perturbation theory on an $8^3$ $\bm{q}$ mesh, using the charge density computed on a $16^3$ $\bm{k}$ mesh. 
The maximally localized Wannier functions are constructed using the \textsc{Wannier90} code~\cite{wannier1,wannier5}. For Al and Pb, four Wannier functions are used to describe the electronic states near the Fermi level. 
These are the $sp^3$-like functions localized along each bond. In the case of Nb, we consider nine Wannier functions with $s$, $p$, and $d$ character. 
In order to solve the anisotropic ME equations, we employ the \textsc{EPW} code~\cite{PhysRevB.76.165108,Margine2013-kc,PONCE2016116,Lee2023-zy}. 
The electronic eigen-energies, phonon frequencies, and electron-phonon
matrix elements are evaluated on $96^3$ $\bm{k}$ and $\bm{q}$ grids. The inner and outer windows are set to $\pm 0.5$~eV and $\pm 15$~eV around the Fermi energy $\varepsilon_{\text{F}}^{(0)}$, respectively. 
A $48^3$ $\bm{k}_{\mathrm{C}}$ grid is used for the summations of the Coulomb contribution in Eqs.~\eqref{phi3}--\eqref{tphi3}. 
The \textit{ab initio} Coulomb parameters $\mu_{\mathrm{C}}=$ 0.251, 0.429, and 0.224 for Al, Nb, and Pb are adopted from Ref.~\onlinecite{Kawamura2020-gr}, where they were obtained with the \textsc{SuperconductingToolkit} (SCTK)~\cite{Kawamura2020-gr}. 
To compute the IR basis functions, we use the \textsc{sparse-ir} library~\cite{Shinaoka2017-cg, Li2020-uz, Wallerberger2023-cc} with $\Lambda=10^6$ and $\epsilon_{\mathrm{IR}}=10^{-8}$, resulting in 96 fermionic and 97 bosonic Matsubara frequency sampling points~\footnote{In practice, we use only 48 sampling frequencies for fermionic functions and 49 frequencies for bosonic functions, as we apply the symmetry relation $G(\im\omega_{j}^{(\mathrm{F})})=G^*(-\im\omega_{j}^{(\mathrm{F})})$}. 

Figure~\ref{fig:2} shows the calculated electron and phonon band structures of Al, Nb, and Pb. 
The electronic band structures and the density of states (DOS) are presented in Figs.~\ref{fig:2}(a), (c), and (e).
The phonon dispersion, the phonon DOS, the isotropic Eliashberg spectral function $\alpha^2 F(\omega)$, and the cumulative electron-phonon coupling strength $\lambda(\omega)$ are presented in Figs.~\ref{fig:2}(b), (d), and (f). 
The band structure of the Wannier model reproduces very well the band structure obtained with the density functional theory (DFT) calculations. 
The computed electron-phonon coupling strength parameters $\lambda$ are found to be 0.432, 1.20, and 1.13 in Al, Nb, and Pb, respectively. These values are comparable with results reported in previous theoretical studies~\cite{Savrasov1996-sd, Liu1996-zc, Akashi2013-nc, Margine2013-kc, PONCE2016116, Kawamura2020-gr, Davydov2020-xo, Pellegrini2023-vr}. 
The electron-phonon coupling strength $\lambda$ of Pb is slightly smaller than the value extracted from tunneling measurements because we neglect the spin-orbit coupling~\cite{Heid2010-nh, PONCE2016116}. \par
In Fig.~\ref{fig:3}, we plot the histograms of the state-dependent electron-phonon coupling strength $\lambda_{n\bm{k}}$ and the superconducting gap function at the lowest Matsubara frequency $\Delta_{n\bm{k}}(\im\pi T)$, denoted as $\rho(\lambda)$ and $\rho(\Delta)$, respectively. 
The two quantities are defined as 
\begin{align}
\rho(\lambda) &= 
\frac{1}{N_{\mathrm{k}}}
\sum_{\bm{k},n}^{\mathrm{Inner.}}
\delta(\lambda_{n\bm{k}}-\lambda)\delta(\varepsilon_{n\bm{k}}-\varepsilon_{\text{F}}^{(0)})
\intertext{with}
\lambda_{n\bm{k}} &= 
\frac{1}{N_{\mathrm{q}}}
\sum_{\bm{q},n'}^{\mathrm{Inner.}}\sum_{\nu}\frac{2\left|g_{n\bm{k},n'{\bm{k}+\bm{q}}}^{\nu}\right|^2}{\omega_{\nu\bm{q}}}\delta(\varepsilon_{n'\bm{k}+\bm{q}}-\varepsilon_{\text{F}}^{(0)})
\end{align}
and
\begin{align}
\rho(\Delta) &= 
\frac{1}{N_{\mathrm{k}}}
\sum_{\bm{k},n}^{\mathrm{Inner.}}
\delta(\Delta_{n\bm{k}}(\im\pi T)-\Delta)\delta(\varepsilon_{n\bm{k}}-\varepsilon_{\text{F}}^{(0)}).
\end{align}
It can be seen that out of the three materials, only Pb displays a two-gap structure, both $\rho(\lambda)$ and $\rho(\Delta)$ histograms being clearly separated into two distinct ranges. 
This two-gap structure has been observed in the differential conductance ($dI/dV(V)$) spectra from tunneling measurements~\cite{Blackford1969-ot, Lykken1971-kc, Ruby2015-mx} and also found in a previous theoretical study based on the density functional theory for superconductors (SCDFT)~\cite{Floris2007-mh}. 
The splitting in the $\rho(\Delta)$ histogram is about 0.25--0.30~meV at 0.2~K, which is slightly larger than the experimentally reported energy separation of 0.15~meV~\cite{Ruby2015-mx}. The transition temperature is estimated by fitting $\Delta(T)\sim A\sqrt{1-T/T_{\mathrm{c}}}$ to the histograms, where $A$ is constant for several temperature points near the transition temperature. 
We obtain $T_{\mathrm{c}}$ values of 2.75, 13.8, and 6.39~K for Al, Nb, and Pb, respectively. 
\par
Here we compare our $T_{\mathrm{c}}$ values with those obtained from literature as shown in Table~\ref{table-tc}. 
Note that the calculations performed in Refs.~\onlinecite{Davydov2020-xo, Pellegrini2023-vr} are isotropic, while those performed in Ref.~\onlinecite{Kawamura2020-gr} are anisotropic. 
Our estimates of $T_{\mathrm{c}}$ for both Al and Nb exceed the experimental values. 
These discrepancies are due to the fact that we have not considered the spin fluctuation in this study~\cite{Kawamura2020-gr, Tsutsumi2020-pc}. 
The estimated value for Al is 2.75~K, relatively high compared to the isotropic calculations in Refs.~\onlinecite{Davydov2020-xo,Pellegrini2023-vr}. 
This could be attributed to the insufficient sampling size of the $\bm{k}_{\mathrm{C}}$ grid. 
Increasing the $\bm{k}_{\mathrm{C}}$-grid sampling to $96^3$ it is estimated that the $T_{\mathrm{c}}$ value would be around 2.2~K, as discussed in Appendix~\ref{sec:conv_tests}.
On the other hand, our $T_{\mathrm{c}}$ value for Nb is comparable to the one obtained from isotropic calculations~\cite{Davydov2020-xo,Pellegrini2023-vr}. 
The close agreement between isotropic and anisotropic calculations implies that the isotropic treatment provides a relatively accurate approximation for Nb, as most of the states exhibit similar values of the gap function. 
In fact, Fig.~\ref{fig:3}(e) shows that the histogram $\rho(\Delta)$ at 0.2~K for Nb has a typical large peak near 2.5~meV. 
For Pb, our calculated $T_{\mathrm{c}}$ is lower than the experimental value. 
This underestimation is associated to the neglect of the spin-orbit interaction in our calculations. 
\par

\begin{table*}
    \centering
  \caption{\label{table-tc} Comparison between the transition temperatures (in Kelvin) obtained in this work and in previous theoretical studies~\cite{Davydov2020-xo, Pellegrini2023-vr, Kawamura2020-gr}. In the table we use the following abbreviations: with spin fluctuation (w SF), without spin fluctuation (w/o SF), with spin-orbit interaction (w SO), and without spin-orbit interaction (w/o SO), respectively. In this work, we employed the anisotropic Eliashberg formalism with a constant Coulomb parameter ($\mu_{\mathrm{C}}$). In the isotropic Eliashberg formalism employed in Refs.~\onlinecite{Davydov2020-xo, Pellegrini2023-vr}, as well as in the isotropic SCDFT formalism employed in Ref.~\onlinecite{Davydov2020-xo}, the Coulomb interaction was treated as an energy-dependent function. The anisotropic SCDFT formalism in Ref.~\onlinecite{Kawamura2020-gr} considered the momentum dependence of the Coulomb interaction. SPG denotes the parameterization proposed by Sanna, Pellegrini, and Gross~\cite{Sanna2020-qv} for the SCDFT formalism. 
  }
  \scalebox{0.87}{
  \begin{tabular}{lccccccccccc}\hline\hline
      \multicolumn{1}{c|}{}&\multicolumn{4}{c|}{A.~Davydov~\textit{et al.}~\cite{Davydov2020-xo}}&\multicolumn{1}{c|}{C.~Pellegrini~\textit{et al.}~\cite{Pellegrini2023-vr}}&\multicolumn{4}{c|}{M.~Kawamura~\textit{et al.}~\cite{Kawamura2020-gr}}&\multicolumn{1}{c|}{This work}&\multicolumn{1}{c}{\multirow{5}{*}{Exp.}}  \\ \cline{1-11}
      \multicolumn{1}{c|}{$\bm{k}$ dep. of $\Delta$}&\multicolumn{4}{c|}{isotropic}&\multicolumn{1}{c|}{isotropic}&\multicolumn{4}{c|}{anisotropic}&\multicolumn{1}{c|}{anisotropic}&\multicolumn{1}{c}{}  \\ \cline{1-11}
      \multicolumn{1}{c|}{formulation}&\multicolumn{2}{c|}{Eliashberg}&\multicolumn{2}{c|}{SCDFT (SPG)}&\multicolumn{1}{c|}{Eliashberg}&\multicolumn{4}{c|}{SCDFT}&\multicolumn{1}{c|}{Eliashberg}&\multicolumn{1}{c}{}  \\ \cline{1-11}
      \multicolumn{1}{c|}{Coulomb int.}&\multicolumn{1}{c|}{Static}&\multicolumn{1}{c|}{Dynamical}&\multicolumn{1}{c|}{Static}&\multicolumn{1}{c|}{Dynamical}&\multicolumn{1}{c|}{Static}&\multicolumn{4}{c|}{Dynamical}&\multicolumn{1}{c|}{Static}&\multicolumn{1}{c}{}  \\ \cline{1-11}
      \multicolumn{1}{c|}{SF \& SO}&\multicolumn{4}{c|}{w/o SF, w/o SO}&\multicolumn{1}{c|}{w/o SF, w/o SO}&\multicolumn{1}{c|}{w/o SF, w/o SO}&\multicolumn{1}{c|}{w SF, w/o SO}&\multicolumn{1}{c|}{w/o SF, w SO}&\multicolumn{1}{c|}{w SF, w SO}&\multicolumn{1}{c|}{w/o SF, w/o SO}&\multicolumn{1}{c}{}  \\ \hline
      Al & 0.9 & 2.5 & 1.6 & 1.3 & 1.03 & 1.9 & 0.89 & 1.9 & 0.88 & 2.75 ($\sim$2.2)\textcolor{blue}{*} & 1.14 \\
      Nb & 13.3 & 23.2 & 7.3 & 7.8 & 12.4 & 14 & 7.6 & 13 & 7.5 & 13.8 & 9.20 \\
      Pb & 6.9 & 8.2 & 5.4 & 3.8 & 6.85 & 4.4 & 3.7 & 6.9 & 6.0 & 6.39 & 7.19 \\ \hline\hline
      \multicolumn{12}{@{}l}{\textcolor{blue}{*} The value in parentheses is a transition temperature roughly estimated from the histogram with the $96^3$ $\bm{k}_{\mathrm{C}}$ grid at 0.2~K as described in Appendix~~\ref{sec:conv_tests}.}\\
  \end{tabular}
  }
\end{table*}

We fit the BCS curve predicted at the weak coupling limit to the histograms, aligning the superconducting gap $\Delta$ at zero temperature with $\Delta(T=0)=1.76\times T_{\mathrm{c}}$ in Figs.~\ref{fig:3}(b), (f), and (j). 
The temperature dependence of the histogram generally agrees well with the BCS curve. 
However, it is evident that the BCS curve does not necessarily pass through the center of the distribution range of the histogram or near the largest peak, particularly in Fig.~\ref{fig:3}(f) for Nb. 
Therefore, accurate estimation of the transition temperature solely by examining the histogram at low temperatures is difficult. \par
Comparing the distribution of $\lambda_{n\bm{k}}$ and that of $\Delta_{n\bm{k}}(\im\pi T)$ on the Fermi surface for each material, it is clear that the momentum distribution of the two quantities are qualitatively similar. 
In the following, we shall primarily focus on discussing the distribution of $\Delta_{n\bm{k}}(\im\pi T)$. 
For Al, there is one large Fermi surface around the $\Gamma$ point and several small Fermi surfaces along each edge around the K points. 
The larger Fermi surface exhibits a range of superconducting gap values between 0.40 and 0.53~meV, whereas the smaller Fermi surfaces have values around 0.35~meV. 
In the calculations for Al, the challenge in achieving convergence with respect to the Brillouin zone sampling most likely arises from the requirement of accurately computing the contribution of these small Fermi surfaces (see also Appendix~\ref{sec:conv_tests}). 
For Nb, there is one large Fermi surface around the $\Gamma$ point and several small Fermi surface on each face around the N point. 
The larger Fermi surface has values around 2.5~meV over almost the entire area, consistent with the fact that the sharp peak in the $\rho(\Delta)$ histogram is centered around 2.5~meV in Fig.~\ref{fig:3}(f). 
The distribution on the smaller Fermi surfaces ranges from 1.9 to 2.4~meV, corresponding to the broadened peak of the histogram $\rho(\Delta)$. 
For Pb, there are two large Fermi surfaces: one is an almost spherical surface encircling the $\Gamma$ point, and the other is a tube-like surface enveloping the edges of the first Brillouin zone. 
The typical value of the gap function varys distinctly for each Fermi surface. 
This separation is also shown in Ref.~\onlinecite{Floris2007-mh} and results in a two-gap structure in the $\rho(\Delta)$ histogram as presented in Fig.~\ref{fig:3}(j). 

\section{Conclusions\label{sec:conc}}

We have presented our implementation of the anisotropic ME formalism coupled with the IR method. 
This approach significantly reduces the computational cost associated with Matsubara frequency sampling, allowing us to account for the Coulomb interaction from electronic states far from the Fermi energy. 
In order to confirm the validity of our methodology, we have conducted calculations for several elemental metals, Al, Nb, and Pb. 
We have found that the estimated transition temperatures are comparable to those obtained from previous first-principles calculations as well as experiments, and the temperature dependence of the gap function is generally consistent with predictions from BCS theory. 
Of particular significance is obtaining the momentum-dependent gap function from the anisotropic ME calculations even at temperatures as low as 0.2~K. 
This quantity can be directly compared with experimental results, and indeed, the gap function for Pb successfully reproduced the two-gap structure observed in experiments. 
We anticipate that our approach will prove instrumental in analyzing superconducting properties at temperatures significantly lower than the transition temperature and potentially estimating other superconducting quantities. 

\acknowledgments
We thank H. Shinaoka for fruitful discussions regarding the utilization of the IR basis functions computed by the \textsc{sparse-ir} library. 
This work was supported by National Science Foundation (Award No. OAC-2103991). 
R.A. was supported by JSPS KAKENHI Grant Number JP24H00190.\par
Calculations were performed on the Frontera supercomputer at the Texas Advanced Computing Center via the Leadership Resource Allocation (LRAC) (allocation DMR22004) and the supercomputer ``MASAMUNE-IMR'' at the Center for Computational Materials Science, Institute for Materials Research, Tohoku University (Project No. 202211-SCKXX-0001 and 202312-SCKXX-0008).

\appendix

\begin{figure*}
  \begin{center}
    \includegraphics[width=17cm]{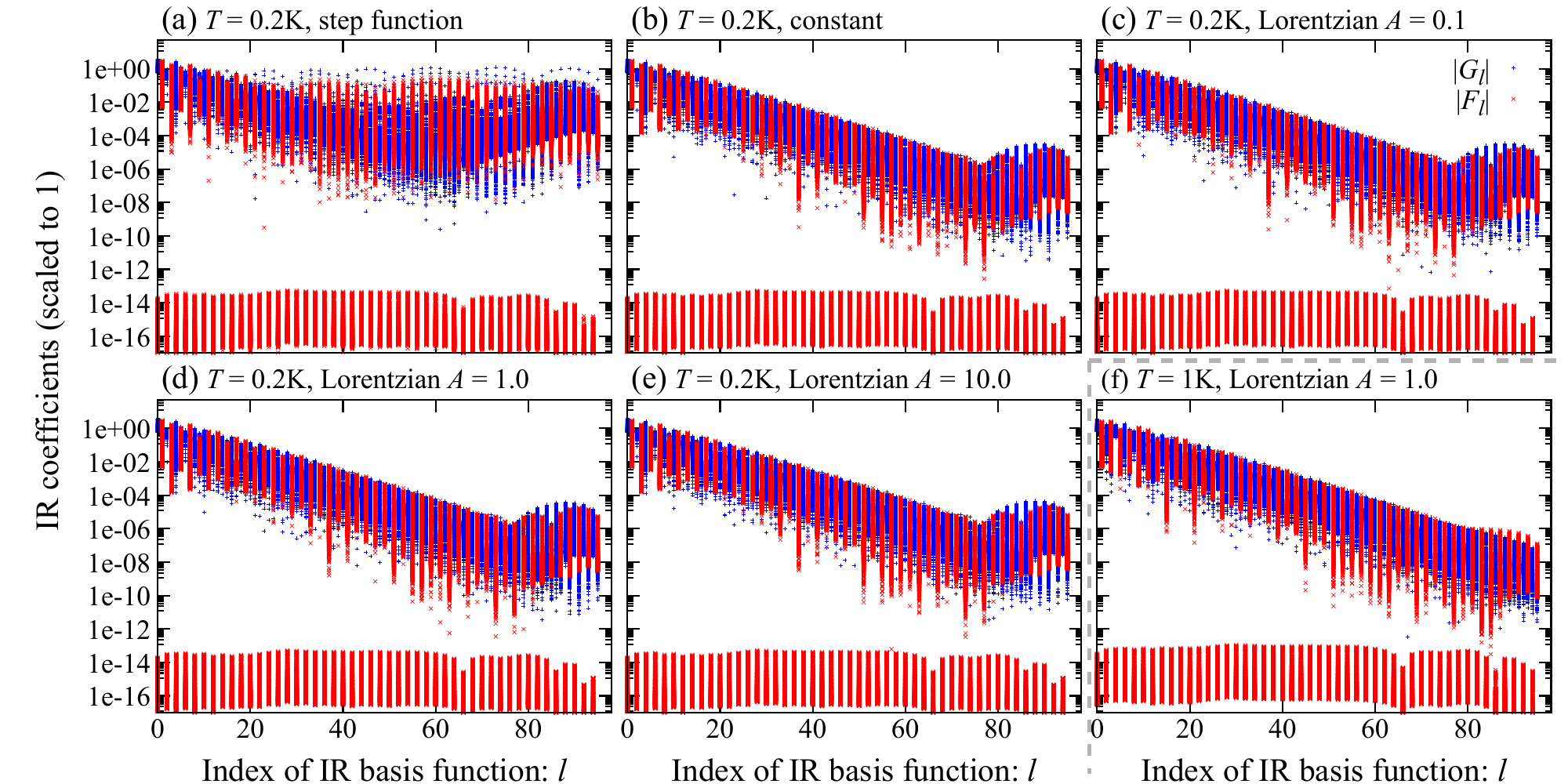}
    \caption{\label{fig:A1}Absolute values of the IR coefficients of both normal and anomalous Green's functions for each index. These coefficients, shown in panels (a)--(e), are extracted from the thirtieth iteration within the self-consistent ME calculation for Nb at $T=0.2$~K. The calculations were conducted using $48^3$ $\bm{k}$ and $\bm{q}$ grids and various initial guesses for the gap function: (a) a step function, (b) a constant, (c) a Lorentzian with $A=0.1$, (d) a Lorentzian with $A=1.0$, and (e) a Lorentzian with $A=10.0$. Panel~(f) shows the same plot as in panel~(d) but the calculation was conducted at $T=1.0$~K. In the ME calculations, the inner window was set to 0.3~eV and the Coulomb interaction was omitted. No noise reduction was applied during the self-consistent procedure. }
 \end{center}
\end{figure*}
\begin{figure*}
  \begin{center}
    \includegraphics[width=17cm]{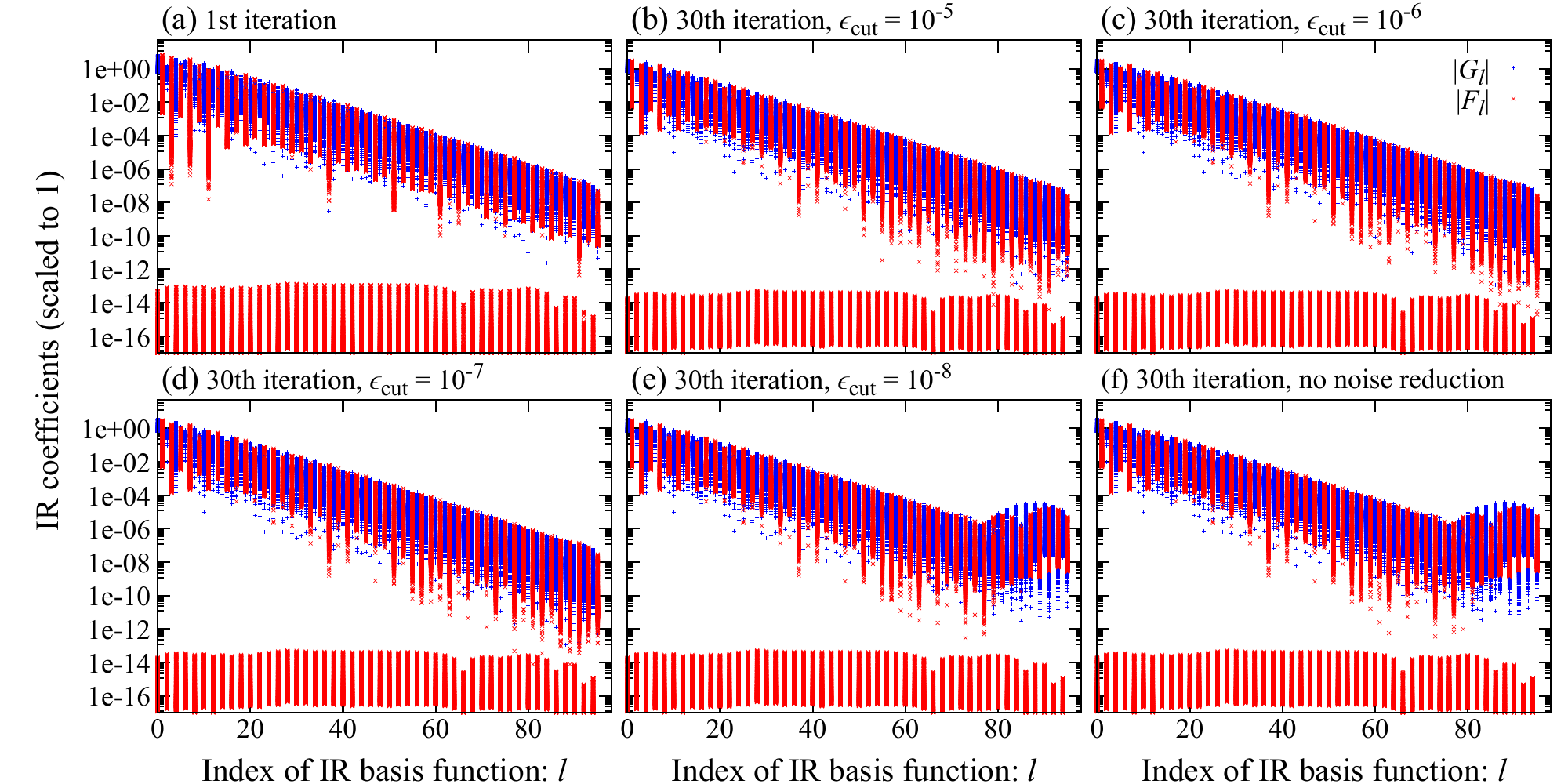}
    \caption{\label{fig:A2}Absolute values of the IR coefficients of both normal and anomalous Green's functions for each index. Panel~(a) shows the coefficients extracted from the first iteration within the self-consistent ME calculation. Panels (b)--(f) show those extracted from the thirtieth iteration immediately after Step (4) in the self-consistent procedure (See Sec.~\ref{sec:calc_details}). In the process of obtaining the coefficients depicted in panels (b)--(f), noise reduction was conducted with varying threshold values for each iteration: (b) $\epsilon_{\mathrm{cut}} = 10^{-5}$, (c) $\epsilon_{\mathrm{cut}} = 10^{-6}$, (d) $\epsilon_{\mathrm{cut}} = 10^{-7}$, and (e) $\epsilon_{\mathrm{cut}} = 10^{-8}$. The coefficients shown in panel~(f) were obtained without applying noise reduction. The ME calculations were conducted for Nb at $T=0.2$~K without the Coulomb interactions using $48^3$ $\bm{k}$ and $\bm{q}$ grids and an inner window of 0.3~eV.}
 \end{center}
\end{figure*}

\section{\texorpdfstring{\uppercase{Decay of the IR coefficients}}{Decay of the IR coefficients}\label{sec:decay}}

When using the IR method, selecting an appropriate function as the initial guess for $\phi_{n\bm{k}}(\im\omega_{j}^{(\mathrm{F})})$ is crucial for successfully fitting the IR coefficients to the Green's function. 
To assess the extent to which the choice of the initial guess influences the fitting of the IR coefficients, we conduct the self-consistent ME calculations for Nb using five different initial guesses. 
The following functions were selected as initial functions: 
\begin{align}
&\phi_{n\bm{k}}(\im\omega_j^{(\mathrm{F})}) 
\notag\\
&\hspace{0.3em}
= 
  \left\{
  \begin{array}{ll} 
    \Delta_0\times H(|\omega_j^{(\mathrm{F})}|-2\omega_{\mathrm{ph}})&\textit{step function}\\
    \Delta_0&\textit{constant}\\
    \displaystyle\frac{\Delta_0}{1+A\times\left(\omega_j^{(\mathrm{F})}/\omega_{\mathrm{ph}}\right)^2}&\textit{Lorentzian}
  \end{array}
  \right.
\end{align}
where $H(x)$ is the Heaviside step function. 
In the test calculations, we examine the effect of sharpness in the Lorentzian by testing $A=0.1$, $1.0$, and $10.0$. 
Figure~\ref{fig:A1} shows the index dependence of the IR coefficients of both normal and anomalous Green's functions with various initial guesses. 
Note that the index $l$ of IR coefficients starts at zero. 
Typically, assuming successful fitting, the IR coefficients should decay rapidly as the index increases~\cite{Shinaoka2017-cg, Chikano2018-qu, Shinaoka2022-iy}. 
In fact, except for panel~(a), a fast decay of the IR coefficients is observed in Fig.~\ref{fig:A1}. 
It can be seen that the width of the Lorentzian does not significantly affect the quality of the fitting, and the fitting is successful even with the constant function. 
In contrast, Fig.~\ref{fig:A1}(a) exhibits no decay of the IR coefficients. 
This indicates that while a step function is not a suitable initial guess for $\phi_{n\bm{k}}(\im\omega_{j}^{(\mathrm{F})})$ in the case of the compact representation using the IR basis functions, both a constant or a Lorentzian function are appropriate. 
Nevertheless, Figs.~\ref{fig:A1}(b)--(e) show that the IR coefficients do not decay completely, instead, they form a small peak in the region where the index $l$ exceeds about 70.
Additionally, while this peak is observed at 0.2~K, it is mostly absent at 1~K as shown in Figs.~\ref{fig:A1}(f). 
To investigate the cause of the peak prominently observed at low temperatures at large indices $l$~\footnote{On rare occasions, the presence of this peak can lead to convergence to a solution with the opposite sign compared to the initial guess.}, we conduct several noise reduction tests using different thresholds in the ME calculations for Nb at $T=0.2$~K. 
In Fig.~\ref{fig:A2}, the peak is not observed at large indices $l$ in the first iteration but appears in the thirtieth iteration in the case when using a lenient threshold or not performing noise reduction. 
As the number of iterations increases, it is observed that the peak becomes larger. 
Notably, when using a threshold higher than $10^{-7}$, the peak does not appear even in the thirtieth iteration. 
It means that the peak observed at large indices $l$ arises from truncation errors accumulated during the iterative procedure in the self-consistent calculation. 
Although the test calculations for Nb using $48^3$ $\bm{k}$ and $\bm{q}$ grids show that $\epsilon_{\mathrm{cut}}=10^{-7}$ is sufficiently large to reduce noise stemming from truncation errors, $\epsilon_{\mathrm{cut}}=10^{-5}$ was applied in other calculations for safety. 
Of course, if one truncates IR coefficients that are not negligible, the accuracy of the restored Green's functions may be compromised. 
It is therefore necessary to use an appropriate value of $\epsilon_{\mathrm{cut}}$.

\section{\texorpdfstring{\uppercase{Special treatment of Matsubara sum convolution}}{Special treatment of Matsubara sum convolution}\label{sec:matsu}} 

The static Coulomb interaction cannot be compactly represented using IR basis functions because a function being constant in the Matsubara frequency domain corresponds to an unbounded spectrum~\cite{Li2020-uz}. 
Therefore, computing a convolution involving the static Coulomb interaction requires a different approach from that of the electron-phonon interaction. 
In this study, we have chosen the gauge such that the imaginary part of the anomalous Green's function is zero, $\mathrm{Im}[F_{n\bm{k}}(\im\omega_j^{(\mathrm{F})})]=0$. This leads to the following relation for the anomalous Green's function, 
\begin{align}
F_{n\bm{k}}(\tau \rightarrow +0)
= T\sum_{j}\mathrm{Re}\left[F_{n\bm{k}}(\im\omega_j^{(\mathrm{F})})\right].
\label{IR-F0}
\end{align}
The convolution with the constant Coulomb interaction $\mu_{\mathrm{C}}/N_{\mathrm{F}}$ appearing in Eqs.~\eqref{phi3}--\eqref{tphi3} can be calculated as follows:
\begin{align}
T\sum_{j'}
F_{n'\bm{k}'}(\im\omega_{j'}^{(\mathrm{F})})
\frac{\mu_{\mathrm{C}}}{N_{\mathrm{F}}}
=
F_{n'\bm{k}'}(\tau \rightarrow +0)\frac{\mu_{\mathrm{C}}}{N_{\mathrm{F}}}.
\end{align}
The Matsubara sum in Eq.~\eqref{ne2} also requires a similar treatment since it is not a convolution with the electron-phonon interaction. 
Regarding the normal Green's function, the following relation holds~\cite{AGDbook1963, Lucrezi2024-bm}: 
\begin{align}
G_{n\bm{k}}(\tau \rightarrow +0)
= -\frac{1}{2}+T\sum_{j}\mathrm{Re}\left[G_{n\bm{k}}(\im\omega_j^{(\mathrm{F})})\right].
\label{IR-G0}
\end{align}
Applying this relation to Eq.~\eqref{ne2} yields the following expression, 
\begin{align}
N_{\e}
=
\frac{2}{N_{\mathrm{k}}}
\sum_{\bm{k},n}
\bigl[
1+
G_{n\bm{k}}(\tau \rightarrow +0)
\bigr].
\label{ne3}
\end{align}
$G_{n\bm{k}}(\tau \rightarrow +0)$ and $F_{n\bm{k}}(\tau \rightarrow +0)$ can be easily evaluated from the IR coefficients. 

\begin{figure*}
  \begin{center}
    \includegraphics[width=17cm]{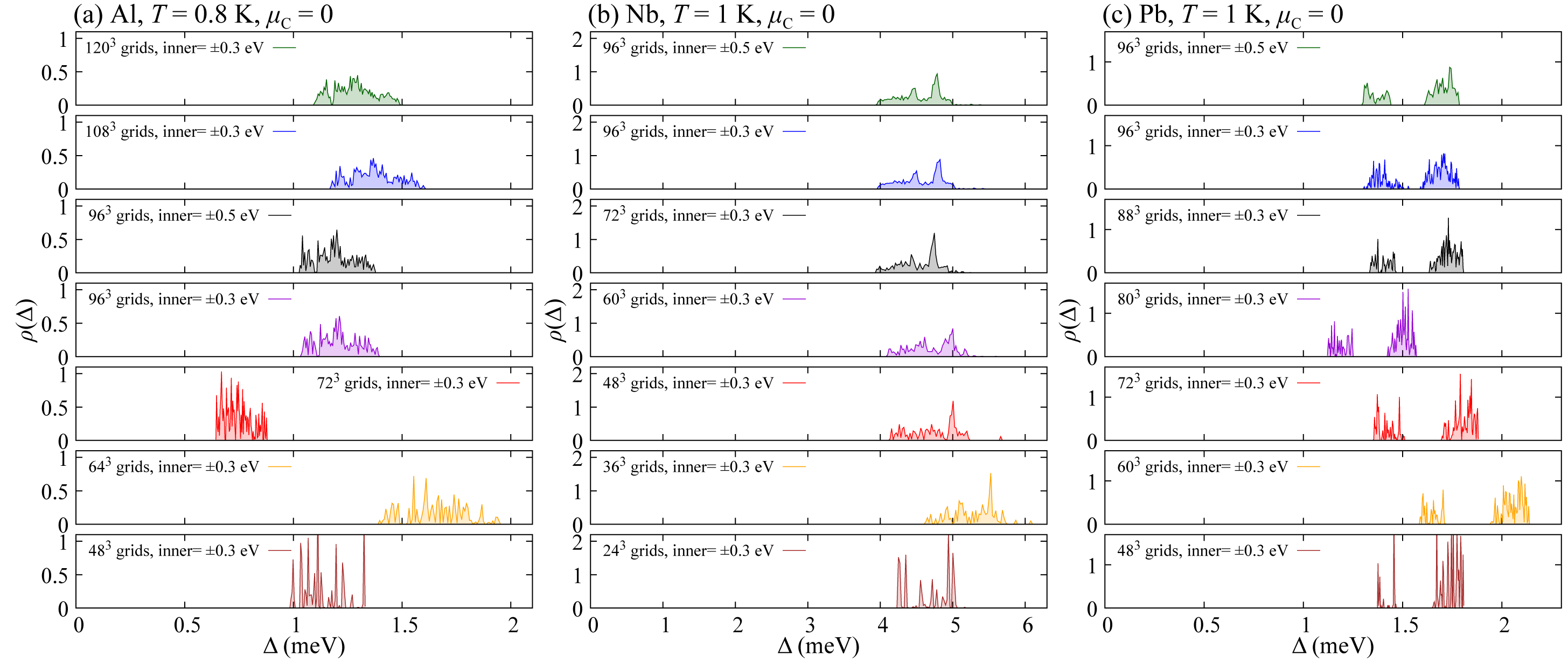}
    \caption{\label{fig:A3}Convergence of the superconducting gap function with respect to the $\bm{k}$- and $\bm{q}$-grid sampling and the inner window. The temperature is set to $T=0.8$~K for Al, and $T=1$~K for Nb and Pb. The calculations were performed without the Coulomb interaction ($\mu_{\mathrm{C}}=0$).}
 \end{center}
\end{figure*}
\begin{figure*}
  \begin{center}
    \includegraphics[width=17cm]{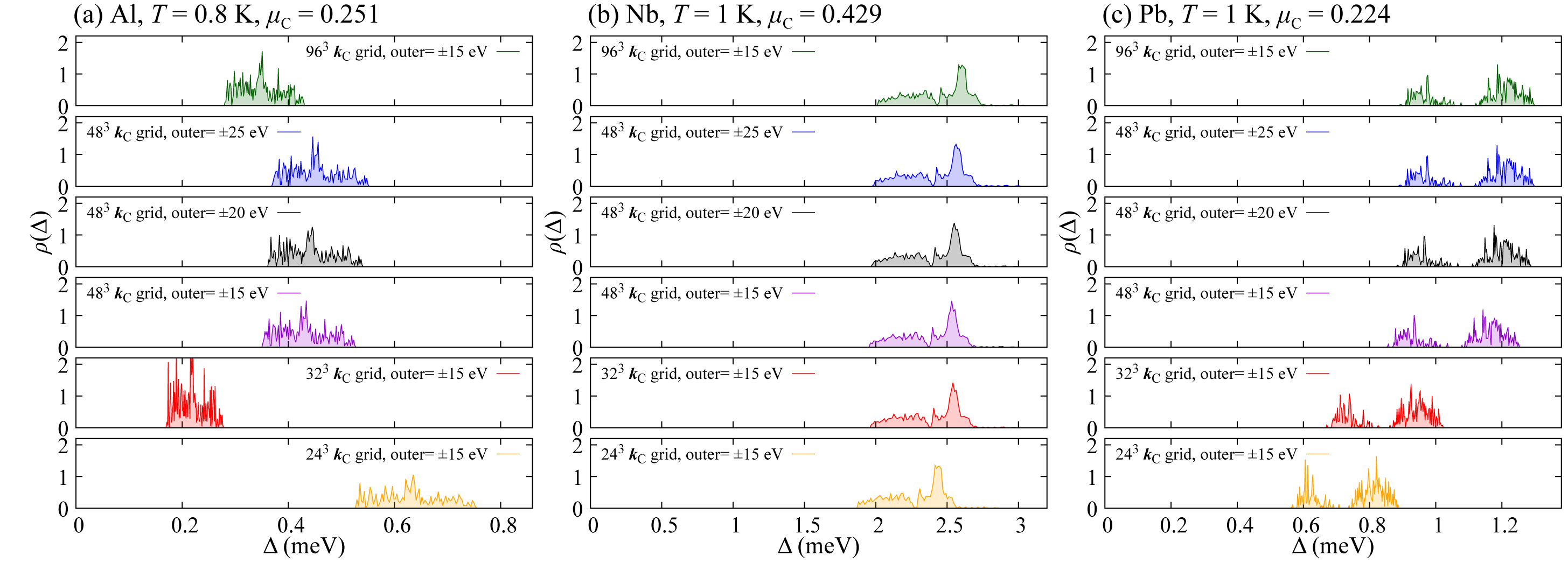}
    \caption{\label{fig:A4}Convergence of the superconducting gap function with respect to the $\bm{k}_{\mathrm{C}}$-grid sampling and the outer window. The calculations were performed with $96^3$ $\bm{k}$ and $\bm{q}$ grids and an inner window of 0.3~eV. The temperature is set to $T=0.8$~K for Al, and $T=1$~K for Nb and Pb. The constant Coulomb parameter is $\mu_{\mathrm{C}}=0.251$ for Al, $\mu_{\mathrm{C}}=0.429$ for Nb, and $\mu_{\mathrm{C}}=0.224$ for Pb.}
 \end{center}
\end{figure*}

\section{\texorpdfstring{\uppercase{Convergence tests}}{Convergence tests}\label{sec:conv_tests}}

\begin{table}[!b]
    \centering
  \caption{\label{table-lambda}Convergence of the electron-phonon couplng strength $\lambda$ with respect to the $\bm{k}$- and $\bm{q}$-grid sampling and the inner window.}
  \begin{tabular}{cccc}\hline\hline
    \multicolumn{1}{c|}{\multirow{2}{*}{system}}&\multicolumn{1}{c|}{sampling size} & \multicolumn{1}{c|}{inner window}& \multicolumn{1}{c}{\multirow{2}{*}{$\lambda$}} \\
    \multicolumn{1}{c|}{}&\multicolumn{1}{c|}{$\bm{k}$ and $\bm{q}$} & \multicolumn{1}{c|}{(eV)}& \\ \hline
    \multirow{9}{*}{Al}&$24^3$--$24^3$ & 0.3 & 0.386 \\
    &$36^3$--$36^3$ & 0.3 & 0.409 \\
    &$48^3$--$48^3$ & 0.3 & 0.358 \\
    &$64^3$--$64^3$ & 0.3 & 0.435 \\
    &$72^3$--$72^3$ & 0.3 & 0.446 \\
    &$96^3$--$96^3$ & 0.3 & 0.432 \\
    &$96^3$--$96^3$ & 0.5 & 0.432 \\
    &$108^3$--$108^3$ & 0.3 & 0.427 \\
    &$120^3$--$120^3$ & 0.3 & 0.426 \\ \hline
    \multirow{9}{*}{Nb}&$24^3$--$24^3$ & 0.3 & 1.14 \\
    &$36^3$--$36^3$ & 0.3 & 1.35 \\
    &$48^3$--$48^3$ & 0.3 & 1.17 \\
    &$60^3$--$60^3$ & 0.3 & 1.22 \\
    &$72^3$--$72^3$ & 0.3 & 1.22 \\
    &$96^3$--$96^3$ & 0.3 & 1.20 \\
    &$96^3$--$96^3$ & 0.5 & 1.20 \\ \hline
    \multirow{9}{*}{Pb}&$24^3$--$24^3$ & 0.3 & 1.65 \\
    &$48^3$--$48^3$ & 0.3 & 1.27 \\
    &$60^3$--$60^3$ & 0.3 & 1.26 \\
    &$72^3$--$72^3$ & 0.3 & 1.20 \\
    &$80^3$--$80^3$ & 0.3 & 1.06 \\
    &$88^3$--$88^3$ & 0.3 & 1.11 \\
    &$96^3$--$96^3$ & 0.3 & 1.13 \\
    &$96^3$--$96^3$ & 0.5 & 1.13 \\ \hline\hline
  \end{tabular}
\end{table}

In this appendix, we provide the results of the convergence tests. Since including the Coulomb contribution would incur high computational costs, we initially conducted the calculations while considering only the electron-phonon interaction term to examine the convergence with respect to the sampling size of the $\bm{k}$ and $\bm{q}$ grids and the size of the inner window. 
The convergence of the electron-phonon coupling strength, $\lambda$, is shown in Table~\ref{table-lambda}, and that of the gap function, $\Delta_{n\bm{k}}(\im\pi T)$, is shown in Fig.~\ref{fig:A3}. 
Note that the gap functions in Fig.~\ref{fig:A3} exhibit larger values compared to those in Fig.~\ref{fig:3} due to the exclusion of the Coulomb term. 
Changing the size of the inner window hardly affects the results, indicating that an inner window of 0.3~eV is sufficient, whereas a size of 0.5~eV was used in the main text. 
Moreover, for Nb and Pb, the calculations converge with $96^3$ $\bm{k}$ and $\bm{q}$ grids. Especially for Nb, acceptable results are obtained even with $48^3$ $\bm{k}$ and $\bm{q}$ grids. 
On the other hand, convergence proves challenging for Al, attributed to the presence of small Fermi surfaces near the edges. Although a $96^3$ $\bm{k}$- and $\bm{q}$-grid sampling is not sufficient in this case, the results are found to be close enough to those obtained for denser grids as shown in Fig.~\ref{fig:A3}(a). 
Considering the balance with the computational cost, the $96^3$ $\bm{k}$- and $\bm{q}$-grid sampling was adopted in the main text for all three metals. 
\par
\begin{figure}
  \begin{center}
    \includegraphics[width=8.6cm]{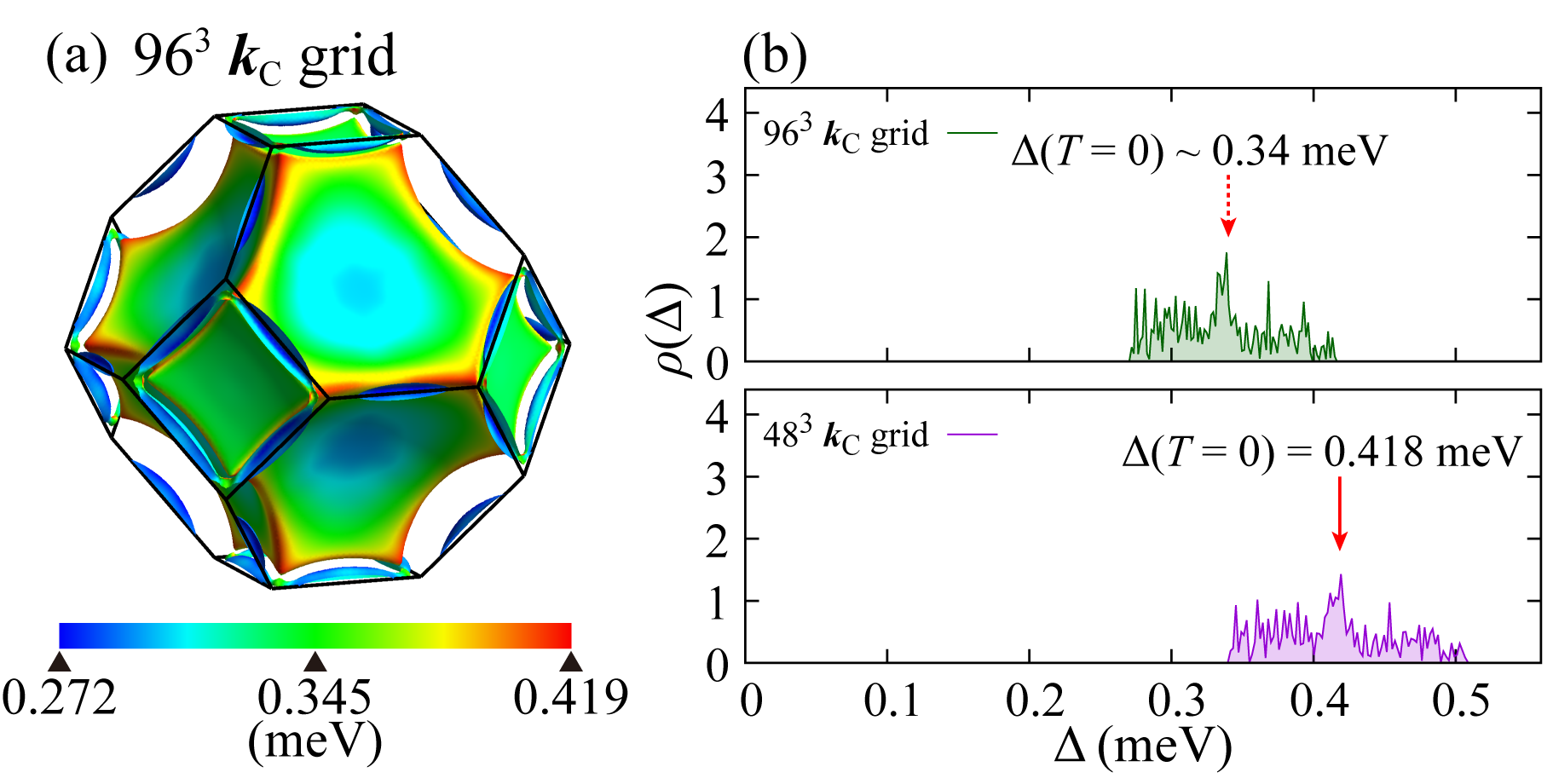}
    \caption{\label{fig:A5}Plots of the superconducting gap function of Al at 0.2~K. The calculations were performed with $96^3$ $\bm{k}$ and $\bm{q}$ grids and an inner window of 0.5~eV: (a) the superconducting gap function with a $96^3$ $\bm{k}_{\mathrm{C}}$ grid on the Fermi surface, and (b) comparison between the histogram of the superconducting gap function with $96^3$ and $48^3$ $\bm{k}_{\mathrm{C}}$ grids. The image on the Fermi surface was rendered using the \textsc{FermiSurfer} software~~\cite{Kawamura2019-ym}.}
 \end{center}
\end{figure}

To investigate the convergence of the gap function with respect to the sampling size of the $\bm{k}_{\mathrm{C}}$ grid and the size of the outer window, we performed calculations by fixing the $\bm{k}$- and $\bm{q}$-grid sampling to $96^3$ and the inner window to 0.3~eV. 
Figure~\ref{fig:A4} demonstrates that as the size of the outer window increases, the gap function gradually increases due to the Coulomb retardation effect arising from the high-energy states. 
However, since the change is small when going from 15 to 25~meV, an outer window of 15~eV was adopted in the main text. 
It is also found that while a $\bm{k}_{\mathrm{C}}$ grid of $32^3$ for Nb and $48^3$ for Pb are sufficient, a $\bm{k}_{\mathrm{C}}$ grid denser than $48^3$ should in principle be used for Al. However, computing the gap function at 0.2~K for Al takes 59,000 CPU core hours using a $96^3$ $\bm{k}_{\mathrm{C}}$ grid, compared to 1,200 using a $48^3$ $\bm{k}_{\mathrm{C}}$ grid. Attempting to compute the full temperature dependence of the gap function with the $96^3$ $\bm{k}_{\mathrm{C}}$ grid, the computational cost is estimated to be about one million core hours. 
While this cost is not unfeasible, it is significantly more expensive, and considering that the momentum dependence of the gap function does not qualitatively change when using the denser grid as shown in Fig.~\ref{fig:A5}(a), we opted to use the $48^3$ $\bm{k}_{\mathrm{C}}$ grid in the main text for Al as well. 
Nevertheless, the histogram of the gap function in Fig.~\ref{fig:A5}(b) allows us to obtain a rough estimate of the transition temperature for the $96^3$ $\bm{k}_{\mathrm{C}}$ grid.
From the BCS curve fitted to the temperature dependence of the gap function for the $48^3$ $\bm{k}_{\mathrm{C}}$ grid, we obtain $\Delta(T=0)=0.418$~meV. Assuming that the BCS curve for the $96^3$ $\bm{k}_{\mathrm{C}}$ grid passes through the same peak of the histogram as in the case of the $48^3$ $\bm{k}_{\mathrm{C}}$ grid, we get $\Delta(T=0)\sim 0.34$~meV, which corresponds to $T_{\mathrm{c}}\sim 2.2$~K. 

\FloatBarrier

\end{document}